\documentclass[twocolumn,aps,prl,showpacs,preprintnumbers, superscriptaddress, amsmath,amssymb]{revtex4}

\usepackage{amsmath}
\usepackage{amsfonts}
\usepackage[all]{xy}
\usepackage{amssymb}
\usepackage{graphicx}%

\usepackage{tabularx}
\usepackage{tikz}

\newcolumntype{M}{>{\centering\arraybackslash}m{1cm}}

\begin{document}

\author{Renan Cabrera}
\email{rcabrera@princeton.edu}
\affiliation{Department of Chemistry, Princeton University, Princeton, NJ 08544, USA} 

\author{Andre G. Campos}
\affiliation{Department of Chemistry, Princeton University, Princeton, NJ 08544, USA}

\author{Denys I. Bondar}
\affiliation{Department of Chemistry, Princeton University, Princeton, NJ 08544, USA}

\author{Herschel A. Rabitz}
\affiliation{Department of Chemistry, Princeton University, Princeton, NJ 08544, USA} 

\title{Dirac open quantum system dynamics: formulations and simulations }

\date{\today}

\begin{abstract}
  We present an open system interaction formalism for the Dirac equation. 
  Overcoming a complexity bottleneck of alternative formulations, our framework
  enables efficient numerical simulations (utilizing a typical desktop) 
  of relativistic dynamics within the  von Neumann 
  density matrix and Wigner phase space descriptions.   
  Employing these instruments, we gain important insights into the effect of quantum dephasing 
  for relativistic systems in many branches of physics.
  In particular, the conditions for robustness of Majorana spinors
  against dephasing are established. 
  Using the Klein paradox and tunneling as examples, we show that 
  quantum dephasing does not suppress negative energy particle generation.
  Hence, the Klein dynamics is also robust to dephasing.

\end{abstract}

\pacs{03.65.Pm, 05.60.Gg, 05.20.Dd, 52.65.Ff, 03.50.Kk}

\maketitle

\section{Introduction.} 

The Dirac equation is a cornerstone of relativistic quantum mechanics
\cite{greiner1990relativistic}. It was originally developed to
describe spin $1/2$ charged particles playing an essential role in the
field of high energy physics
\cite{elze1986transport,Hakim2011introduction,zee2010quantum}.
Recently, there is resurging interest in the Dirac equation
because it was found to be an effective dynamical model of
unexpectedly diverse phenomena occurring in high-intensity lasers
\cite{RevModPhys.84.1177}, solid state
\cite{novoselov2005two,katsnelson2006chiral,hasan2010colloquium,zhang2015perfect},
optics \cite{otterbach2009confining,ahrens2015simulation}, cold atoms
\cite{vaishnav2008observing,boada2011dirac}, trapped ions
\cite{gerritsma2010quantum,blatt2012quantum}, circuit QED
\cite{pedernales2013quantum}, and the chemistry of heavy elements
\cite{C1CP21718F,autschbach2012perspective}. However, there is a need
to go beyond coherent dynamics offered by the Dirac equation alone in
order to model the effects of imperfections, noise, and interaction
with a thermal bath \cite{nandkishore2014rare}. To construct such
models, we will first review how these effects are described
without relativistic considerations \cite{gardiner2004quantum}.

In the non-relativistic regime, the Schr\"odinger equation describes a
quantum systems isolated from the rest of the universe. This is a good
approximation for certain conditions. For example, an atom in a dilute
gas can be considered to be a closed system if the time scale of
the dynamics is much faster than the mean collision time. If we
would like to include collisions in the picture, we need to keep
track of the quantum phases of each atom in the gas. This is
unfeasible. This type of dynamics motivated development of the theory
of \emph{open quantum systems} \cite{petruccione2002theory}, where a
single particle picture is retained albeit with more general dynamical
equations. There are two methods to introduce interactions with an
environment: (i) the Schr\"odinger equation with an additional
stochastic force, or (ii) the conceptually different density matrix
formalism \cite{gardiner2004quantum}. In the latter, a state of an
open quantum system is represented by a self-adjoint density operator
$\hat{\rho}$ with non-negative eigenvalues summing up to one. The
master equation, governing evolution of $\hat{\rho}$, reads
\begin{equation}\label{eq1}
	i \hbar \frac{d}{dt} \hat{\rho} = [ \hat{H}, \hat{\rho}] + \mathcal{D}( \hat{\rho} ) ,
\end{equation}
where $\hat{H}$ is the quantum Hamiltonian and the dissipator
$\mathcal{D}( \hat{\rho} )$ encodes the interaction with an
environment. The von Neumann equation \cite{gardiner2004quantum}
describing unitary evolution is recovered by ignoring the
dissipator. When $\mathcal{D}( \hat{\rho} ) \neq 0$, Eq. (\ref{eq1})
generally does not preserve the von Neumann entropy $S=- {\rm Tr}\,
(\hat{\rho} \log \hat{\rho} )$, which measures the amount of information
stored in a quantum system. We note that effective elimination of $\mathcal{D}(
\hat{\rho} )$ is a fundamental challenge in order to develop 
many quantum  technologies \cite{sarandy2005adiabatic,viola1999dynamical}.

The non-relativistic theory of open quantum systems provided profound
insights into some fundamental questions of physics such as the
emergence of the classical world from the quantum one
\cite{RevModPhys.75.715,zurek1991decoherence,adler2007collapse,PhysRevLett.89.170405,1402-4896-1998-T76-027,PhysRevLett.88.040402,PhysRevLett.96.010403},
measurement theory
\cite{RevModPhys.75.715,PhysRevA.67.042103,jacobs2006straightforward,zurek2009quantum},
quantum chaos
\cite{zurek2001sub,PhysRevLett.89.170405,PhysRevLett.96.010403} and synchrotron radiation 
\cite{Bazarov2012SynchrotronWigner,gasbarro2014reducedWigner,Tanaka2014SynchrotonWigner}.

To study the quantum-to-classical transition, it is instrumental to
put both mechanics on the same mathematical footing
\cite{blokhintsev2010philosophy,
  Heller1976,PhysRevLett.80.4361,zurek1991decoherence,1402-4896-1998-T76-027,RevModPhys.75.715,PhysRevA.67.042103,bolivar2004quantum,
  zachos2005quantum,
  Kapral2006,PhysRevLett.109.190403,Polkovnikov2010}. This is achieved
by the Wigner quasi-probability distribution $W(x,p)$
\cite{Wigner1932}, which is a phase-space representation of the
density operator $\hat{\rho}$. Note that the Wigner function serves as
a basis for a self-consistent phase space representation of quantum
mechanics \cite{zachos2005quantum, Curtright2013}, which is equivalent
to the density matrix formalism.

Previous attempts to construct the relativistic theory of open quantum
system relied on the relativistic extension of the Wigner function
without introducing the corresponding density matrix formalism. In
Sec. \ref{Sec:GeneralForm}, we will first present the manifestly
covariant density matrix formalism for a Dirac particle and then
construct the Wigner representation. The development of the
relativistic Wigner function was motivated by applications in quantum
plasma dynamics and relativistic statistical mechanics
\cite{Hakim2011introduction}.  The manifestly covariant relativistic
Wigner formalism for the Dirac equation was put forth in
Refs. \cite{hakim1978covariant,hakim1982,elze1986transport,vasak1987quantum}
(see Ref. \cite{Hakim2011introduction} for a comprehensive review). In
addition, exact solutions for physically relevant systems were
reported in Refs. \cite{yuan2010wigner,kai2011wigner}. The following
conceptual difference between the non-relativistic and relativistic
Wigner functions was elucidated in Ref. \cite{campos2014violation}: In
non-relativistic dynamics, Hudson's theorem states that the Wigner function for a
pure state is positive if and only if the underlying wave function is
a Gaussian \cite{Hudson1974249}. In other cases, the Wigner function
contains negative values. However, this statement does not carry over
to the relativistic regime. In particular, there are many physically
meaningful spinors whose Wigner function is positive
\cite{campos2014violation}. Note that the Wigner function's negativity
is an important resource in quantum information theory \cite{emerson,
  mari}.

The limit $\hbar \rightarrow 0$ of the non-relativistic Wigner
function is non-singular and recovers classical mechanics. The same
limiting property is expected from the relativistic
extension. However, the manifest covariance of the relativistic Wigner
function needed to be broken in order to perform the $\hbar
\rightarrow 0$ limit \cite{Bialynicki-birula1977,vasak1987quantum,
  Shin1993}. From a different perspective, the covariant classical
limit was obtained in
Refs. \cite{bolivar2001classical,bolivar2004quantum}. In Appendix
\ref{Sec:ClassLimitDirac} of the current work, we provide a simpler
manifestly-covariant derivation of the classical limit. Contrary to
the previous work, our derivation recovers two decoupled classical
equations of motion: one governing the dynamics of positive energy
particles and the other describing negative energy particles (i.e.,
antiparticles). This classical limit of the Dirac equation is an example 
of classical Nambu dynamics \cite{nambu1973generalized}.

An alternative quantum field theoretic formulation of the Wigner function for Dirac fermions has also been put forth \cite{Bialynicki-birula1977, Bialynicki-Birula1991,shin1992wigner, bialynicki2014relativistic, hebenstreit2011strong, berenyi2014describing, blinne2016comparison}. 

As mentioned before, the current interest in the Dirac equation goes
far beyond relativistic physics. These new opportunities come along
with new challenges. It is the aim of the current Article to overcome
some of those problems by furnishing a new formulation of traditional
(i.e., closed system) relativistic dynamics enabling efficient
numerical simulations as well as physically consistent inclusion of
open system interactions. We believe that the developed formalism and
numerical methods will influence the following fields:
\begin{enumerate}

\item \emph{Understanding the role of the environment for the
    classical world emergence.}  In particular, we elucidate the
  influence of decoherence (i.e., loss of quantum phase coherence) on
  relativistic dynamics in Secs. \ref{Sec:MajoranaIlustrations} and
  \ref{Sec:KleinIllustrations}, where Klein tunneling
  \cite{katsnelson2006chiral} and the associated paradox are analyzed along with
  the Majorana fermion dynamics.

 \item \emph{Development of the quantum relativistic theory of energy
     dissipation.}  Based on existing models of non-relativistic
   quantum friction \cite{kohen1997phase,Bondar-WignerLindblad-2016}, we expect a
   relativistic model of energy damping to obey: (i) the mass-shell
   constraint, (ii) translational invariance (in particular, the dynamics
   should not depend on the choice of the origin), (iii) equilibration
   (the model should reach a steady state at long time propagation. In
   particular, the final energy at $t\to +\infty$ should be bounded
   thereby preventing runaway population of the negative energy
   continuum), (iv) thermalization (i.e., the achieved steady state should
   represent  thermal equilibrium), (v) relativistic extension of
   Ehrenfest theorems (i.e., see the dynamical constraints for expectation values
   encompassing energy drain in Ref. \cite{Bondar-WignerLindblad-2016}). Some
   preliminary steps towards the desired relativistic model are
   reported in Ref. \cite{Campos2016}.
 
 \item \emph{Modeling environmental effects in Dirac materials} such
   as topological insulators
   \cite{hasan2010colloquium,bernevig2013topological,wang2016hourglass},
   Weyl semimetals \cite{lv2015experimental,inoue2016quasiparticle},
   and graphene \cite{novoselov2005two}. In these cases, open system
   dynamics models sample impurities and imperfections as well as
   external noise. Recently, the Dirac equation with an additional
   stochastic force was utilized for this purpose
   \cite{nandkishore2014rare}. To the best of our knowledge, a more
   general master equation formalism is yet to be explored.

\item \emph{Understanding robustness of a Majorana particle}, which is
  defined as being its own antiparticle.  Experimental implementation
  of solid-state analogues of Majorana fermions
  \cite{alicea2012new,buhler2014majorana,nadj2014observation} opens up
  possibilities to study the physics of these unusual states. In
  particular, Majorana bound states are well suited components of
  topological quantum computers \cite{RevModPhys.80.1083}. Due to its
  topological nature, Majorana states are expected to be robust
  against perturbations and imperfections
  \cite{albrecht2016exponential}. Dissipative dynamics modeled within
  a Lindblad master equation confirmed a significant degree of
  robustness in a specific optical lattice
  \cite{diehl2011topology}. However, the robustness is not universal
  \cite{budich2012failure} and there is a need for enhancement (e.g.,
  employing error correction techniques \cite{PhysRevX.4.011051}).
  Note that Majorana states studied in condensed matter physics
  \cite{alicea2012new,buhler2014majorana,nadj2014observation},
  do not strictly coincide with the authentic Majorana
  spinors \cite{majorana1937teoria}, albeit sharing common
  features. In the present paper, we consider original Dirac
  Majorana spinors \cite{majorana1937teoria}. In
  Sec. \ref{Sec:MajoranaIlustrations}, we demonstrate that a
  single-particle Majorana spinor exhibits robustness even for strong
  couplings to the dephasing environment, which otherwise quickly
  washes out interferences for particle-particle superpositions (aka,
  Schr\"{o}dinger cat states). Moreover, this phenomenon has an
  intuitive explanation in the phase-space representation, where quantum dephasing turned out to be equivalent 
  to Gaussian filtering over the momentum axis (detailed explanation in Secs. \ref{Sec:OpenSysInter} and \ref{Sec:NumAlg}). The
  applicability of this insight to condensed matter systems should be a
  subject of further studies.
 
\item \emph{Development of manifestly covariant quantum open system
    interaction.}  Coupling a Dirac particle to the environment
  generally introduces a preferred frame of reference, thereby
  breaking the Lorentz invariance.  However, coupling to the vacuum,
  causing spontaneous emission, Lamb shift \emph{etc.}
  \cite{Welton1948}, and radiation reaction
  \cite{Baylis20027,PhysRevD.88.025021}, needs to be manifestly
  covariant because the vacuum has no preferred frame of reference. Solid
  state physics holds a promise to implement many exotic quantum
  effects experimentally not yet verified \cite{eisert2015quantum},
  e.g., the Unruh effect and Hawking radiation. Solid state dynamics
  naturally includes the interaction with the environment, thus the need to
  include open system interaction into the dynamics of interest. A
  relativistic quantum theory of measurements also requires
  development of manifestly covariant master equations. Currently,
  approaches based on axiomatics \cite{sorkin1995quantum}, stochastic
  Dirac and Lindblad master equations \cite{breuer2000relativistic}
  are explored.  Nevertheless, the proposed equations are
  computationally unfeasible at present. In the current work, we lay
  the ground for a computationally efficient technique by introducing
  a manifestly covariant von Neumann equation (see
  Sec. \ref{Sec:GeneralForm}) based on
  Refs. \cite{hakim1978covariant,hakim1982,elze1986transport,vasak1987quantum,Hakim2011introduction}.
\end{enumerate}

This paper is organized in seven sections and two appendices. Section
\ref{Sec:GeneralForm} provides the general mathematical formalism
including the manifestly relativistic covariant von Neumann
equation. Section \ref{Sec:RelativisticWigner} is concerned with the
relativistic Wigner function and related representations. Section
\ref{Sec:OpenSysInter} introduces open system interactions by
considering a model of dephasing, environmental interaction leading to
the loss of quantum phase.  Numerical algorithms are developed in
Sec. \ref{Sec:NumAlg} and illustrated for the dynamics of Majorana
spinors and the Klein paradox in Secs. \ref{Sec:MajoranaIlustrations} and
\ref{Sec:KleinIllustrations}, respectively. The final section \ref{Sec:conclusions}  provides
the conclusions. Appendix \ref{Sec:LorentzCovariance} treats the
concept of relativistic covariance, and Appendix
\ref{Sec:ClassLimitDirac} elaborates the classical limit ($\hbar
\rightarrow 0$) of the Dirac equation in manifestly covariant fashion.

%%%%%%%%%%%%%%%%%%%%%%%%%

\section{ General Formalism }\label{Sec:GeneralForm}

Note that throughout the paper,  $\boldsymbol{x}$
and $x$ denote different variables; likewise, $ \hat{\boldsymbol{x}}$ and $\hat{x}$ denote different operators. In addition,
Greek characters (e.g., $\mu$, $\nu$), used as indices for Minkowski vectors,
are assumed to run from $0$ to $3$; while, Latin indices (e.g., $j$, $k$) run from $1$
to $3$. The Minkowski metric is a diagonal matrix 
${\rm diag}(1,-1,-1,-1)$. This implies that $x^0 = x_0$ and $x^k = -x_k$.

The manifestly covariant Dirac equation reads
\begin{align}
\label{Dirac-Eq}
 D(  \hat{\boldsymbol{x}}^{\mu}  , \hat{\boldsymbol{p}}_{\mu} ) | \psi \rangle &= 0 ,
\end{align}
where the Dirac generator $ D(  \hat{\boldsymbol{x}}^{\mu}  , \hat{\boldsymbol{p}}_{\mu} )$  and the commutation 
relations are defined as
\begin{align}
 D( \hat{\boldsymbol{x}}^{\mu} , \hat{\boldsymbol{p}}_{\mu} )  
  &= \gamma^{\mu} [ c \hat{\boldsymbol{p}}_{\mu} - eA_{\mu}(  \hat{\boldsymbol{x}}  ) ] - mc^2, \\
  {[}  \hat{\boldsymbol{x}}^{\mu}  ,   \hat{\boldsymbol{p}}_{\nu}  {]}  &= -i \hbar \delta^{\mu}_{\,\,\, \nu}.  \label{XPCommutator}
\end{align}
Note that the negative sign in the right hand side of Eq. (\ref{XPCommutator}) occurs due to the fact  
\begin{align}
 {[}  \hat{\boldsymbol{x}}^{k}  ,   \hat{\boldsymbol{p}}_{j}  {]}  = -i \hbar \delta^{k}_{\,\,\, j}  
  \longleftrightarrow 
 {[}  \hat{\boldsymbol{x}}^{k}  ,   \hat{\boldsymbol{p}}^{j}  {]}  = i \hbar \delta^{k j},  
\end{align}
in agreement with non-relativistic dynamics where the momentum is expressed in contravariant components $\hat{\boldsymbol{p}}^{j}$. 

From the well established work on relativistic 
statistical quantum mechanics \cite{hakim1978covariant,hakim1982,elze1986transport,vasak1987quantum,Hakim2011introduction}, 
the manifestly covariant von Neumann equation can be written as
\begin{align}
 \label{relativistic-vonNeumann}
      D( \hat{\boldsymbol{x}}^{\mu} , \hat{\boldsymbol{p}}^{\mu}  ) \hat{P} = 0, \qquad
  \hat{P}  D( \hat{\boldsymbol{x}}^{\mu} , \hat{\boldsymbol{p}}^{\mu}  )   = 0,
\end{align}
where $\hat{P}$ represents the density state operator acting on 
the \emph{ {\bf M}anifestly {\bf C}ovariant {\bf S}pinorial Hilbert space} (MCS). 
Equation  (\ref{relativistic-vonNeumann}) is the foundation for all 
the subsequent developments.

Following Ref. \cite{bondar2012wigner,PhysRevA.92.042122}, we
introduce the \emph{{\bf M}anifestly {\bf C}ovariant Hilbert {\bf P}hase space} (MCP) where the algebra of observables 
consists of  $( {\hat{\boldsymbol{x}}},{\hat{\boldsymbol{p}}_{\mu} }  )$ [see Eq. (\ref{XPCommutator})] along with the mirror operators  $( {\hat{\boldsymbol{x}}^{\prime \mu }} ,  {\hat{\boldsymbol{p}}_{\mu}^{\prime} }  )$ obeying
\begin{align}
  {[}  \hat{\boldsymbol{x}}^{\mu}  ,   \hat{\boldsymbol{p}}_{\nu}  {]}  = -i \hbar \delta^{\mu}_{\,\,\, \nu}, \qquad
 {[}  {\hat{\boldsymbol{x}}^{\prime \mu }}  ,   \hat{\boldsymbol{p}}_{\nu}^{\prime}  {]}  = i \hbar \delta^{\mu}_{\,\,\, \nu}, 
\label{commutation-mirrow}
\end{align}
and all the other commutators vanish.
In MCP the role of density operator  $\hat{P}$ is taken over by the ket state $|P\rangle$ according to
\begin{align}
\label{rule1}
  \hat{O}(  \hat{\boldsymbol{x}}^{\mu} , \hat{\boldsymbol{p}}^{\mu}  )  \hat{P}
 \quad \longleftrightarrow \quad & \, 
\overrightarrow{O}(  \hat{\boldsymbol{x}}^{\mu} , \hat{\boldsymbol{p}}^{\mu}  ) | P \rangle, \\
\label{rule2} 
 \hat{P}  \hat{O}(  \hat{\boldsymbol{x}}^{\mu} , \hat{\boldsymbol{p}}^{\mu}  )  
 \quad \longleftrightarrow  \quad &\,
   | P \rangle   \overleftarrow{O} 
(\,  \hat{\boldsymbol{x}}^{\prime \mu}  ,\,   \hat{\boldsymbol{p}}^{\prime \mu}   \,),
  \end{align}
where the arrows indicate the direction of application of the 
operators $O(  \hat{\boldsymbol{x}}^{\mu} , \hat{\boldsymbol{p}}^{\mu}  )$ and  $O(\,  \hat{\boldsymbol{x}}^{\prime \mu}  ,\,   \hat{\boldsymbol{p}}^{\prime \mu}   \,)$.
Thus, the relativistic von Neumann equation (\ref{relativistic-vonNeumann}) reads in MCP as
\begin{align}
  \label{MCP-vonNeumann}
      \overrightarrow{D}( \hat{\boldsymbol{x}}^{\mu} , \hat{\boldsymbol{p}}^{\mu}  ) | P \rangle = 0, \qquad
   | P \rangle   \overleftarrow{D} 
(\,  \hat{\boldsymbol{x}}^{\prime \mu}  ,\,  \hat{\boldsymbol{p}}^{\prime \mu}   \,) = 0.
\end{align}
A summary of the two introduced formulations is given in
Table \ref{tab-Manifestly}. 
\begin{table}
  
  \begin{tabularx}{0.7\textwidth}{|c|c|c|}
   \cline{1-3}
    &    &   \\   
    &    \begin{tabular}{c}
     {\bf M}anifestly {\bf C}ovariant \\ {\bf S}pinorial Hilbert space \\ MCS
        \end{tabular} 
        & \begin{tabular}{c} {\bf M}anifestly {\bf C}ovariant  \\ Hilbert {\bf P}hase space \\ MCP \end{tabular}  \\  
    &    &    \\   \cline{1-3}
    &    &   \\ 
 State &$\hat{P}$  & $|P\rangle$   \\
   &     &   \\ \cline{1-3}
   &     & \\
 Operators &$ \hat{O}(  \hat{\boldsymbol{x}}^{\mu} , \hat{\boldsymbol{p}}^{\mu}  )$ &
  $\overrightarrow{O}(  \hat{\boldsymbol{x}}^{\mu} , \hat{\boldsymbol{p}}^{\mu}  )$, 
  $ \overleftarrow{O} (  \hat{\boldsymbol{x}}^{\prime \mu}  ,   \hat{\boldsymbol{p}}^{\prime \mu}   )$ \\
    &   & \\
  &&\\ \cline{1-3}
    &    &  \\
 Equation   &   
      $D( \hat{\boldsymbol{x}}^{\mu} , \hat{\boldsymbol{p}}^{\mu}  ) \hat{P} = 0$       &
      $ \overrightarrow{D}( \hat{\boldsymbol{x}}^{\mu} , \hat{\boldsymbol{p}}^{\mu}  ) | P \rangle = 0$ \\
 of motion &&\\
      & $\hat{P}  D( \hat{\boldsymbol{x}}^{\mu} , \hat{\boldsymbol{p}}^{\mu}  )   = 0 $ &   
      $ | P \rangle   \overleftarrow{D} (\,  \hat{\boldsymbol{x}}^{\prime \mu}  ,\,  \hat{\boldsymbol{p}}^{\prime \mu}   \,) = 0$ \\
   &   & \\
   \cline{1-3}   
  \end{tabularx}
\caption{ Two  manifestly covariant formulations of relativistic quantum mechanics. }
\label{tab-Manifestly}
\end{table}

The manifest covariance of Eq. (\ref{MCP-vonNeumann}) can
be relaxed to implicit  covariance by separating the
time according to the $3+1$ splitting $\hat{\boldsymbol{x}}^\mu = (
c\hat{\boldsymbol{t}} , \hat{\boldsymbol{x}}^k )$
\cite{alcubierre2008introduction}. This means that the underlying
relativistic covariance is maintained but it is no longer evident. In the spirit of the $3+1$ scheme we define the Dirac
Hamiltonian as
\begin{align} 
\label{Dirac-Hamiltonian}
\hat{H}  = 
    \alpha^k[c \hat{\boldsymbol{p}}^k - e A^k( \hat{\boldsymbol{t}} , \hat{\boldsymbol{x}}^k  )  ] + mc^2 \gamma^0 
 +  e A^0( \hat{\boldsymbol{t}} , \hat{\boldsymbol{x}}^k ).  
\end{align}
The von-Neumann equation (\ref{MCP-vonNeumann})  in 
the \emph{ {\bf I}mplicit {\bf C}ovariant Hilbert {\bf P}hase space} (ICP) becomes
\begin{align}
\label{ICHP-vonNeumann-1}
   \left[   c \overrightarrow{ \hat{\boldsymbol{p}}_{0}} - 
  \overrightarrow{H}( \hat{\boldsymbol{t}} ,  \hat{\boldsymbol{x}}^k  ,  \hat{\boldsymbol{p}}_k  )    \right] |P\rangle \gamma^0 = 0,  \\
  |P \rangle  \gamma^{0}  
   \left[  c \overleftarrow{ \hat{\boldsymbol{p}}_{0}^{\prime}} -  
  \overleftarrow{H}( \hat{\boldsymbol{t}}^\prime ,  \hat{\boldsymbol{x}}^{\prime k}  ,  \hat{\boldsymbol{p}}_{k}^{\prime}  ) 
  \right] = 0.  
\label{ICHP-vonNeumann-2}
\end{align}

\begin{figure}
  \centering
      \includegraphics[width=0.7\hsize]{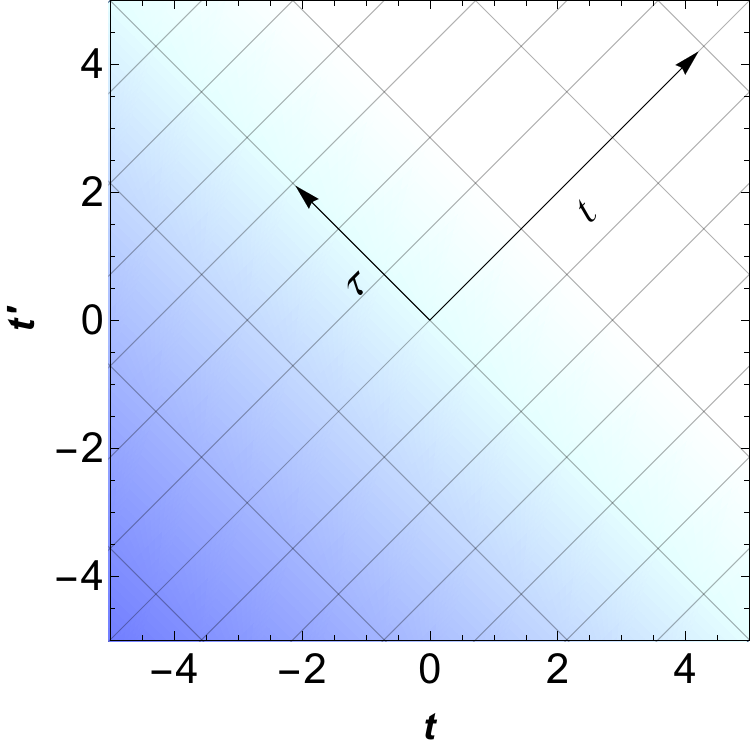}
      \caption{ (Color online) Graphical illustration of the relation between the
        double time variables in the ICP space as defined in Table \ref{tab-BoppOperators}. 
        The color gradient is directed along
        the $t$ coordinate.  }
  \label{fig:time-variables}
\end{figure}

\begin{figure}
  \centering
      \includegraphics[width=.83\hsize]{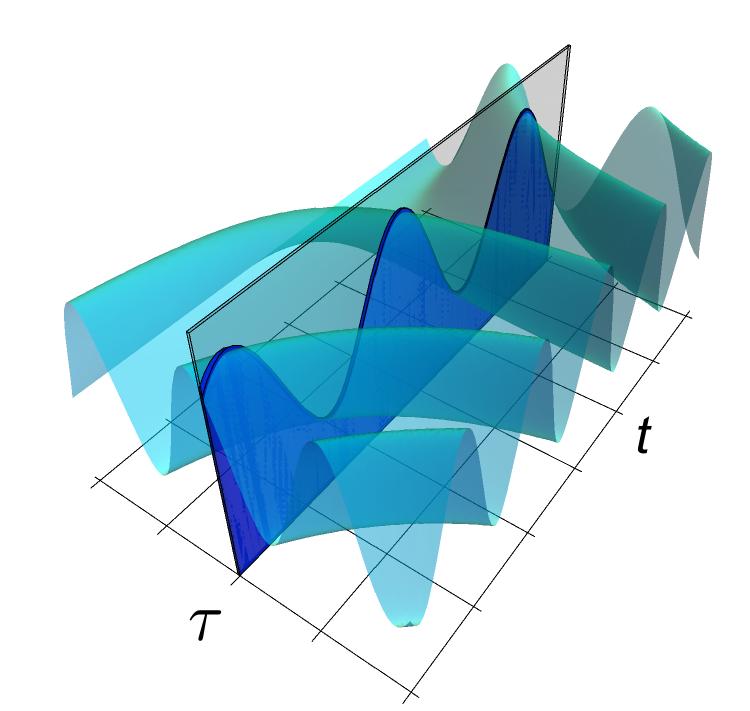}
      \caption{ (Color online) Schematic illustration of a quantum state
        propagating along time $t$ within the slice $\tau=0$ according to
        Eq. (\ref{SCP-vonNeumann}).  A different inertial reference
        frame would generate another slice.   }
  \label{fig:time-slice}
\end{figure}

Inspired by the Bopp transformations in the non-relativistic quantum mechanical phase 
space  \cite{Bopp1956,hillery1984distribution}, a 
 representation of the algebra  (\ref{commutation-mirrow}) can be constructed in terms of ICP Bopp operators 
$( \hat{t}, \hat{\tau}, \hat{\Omega} , \hat{E}, \hat{x}^k , \hat{p}_k , \hat{\lambda}_k , \hat{\theta}^k ) $ in Table \ref{tab-BoppOperators},
\begin{table}
  \centering
  \begin{tabular}{|c|c|c|}
   \cline{1-3}
   &     &   \\   
   &  ICP operators   & Mirror ICP operators  \\  
   &     &    \\   \cline{1-3}
   &     &   \\ 
 \begin{tabular}{c}
%            \\
 Space-time \\
            \\
            \\ \cline{1-1}
            \\
            \\
 Momentum-energy\\
%            \\  
 \end{tabular} 
  &\begin{tabular}{ccc} 
      $\hat{\boldsymbol{t}}$& $=$ & $ \hat{t} - \frac{1}{2} \hat{\tau}$    \\
        &   &  \\
    $\hat{\boldsymbol{x}}^k$&$=$& $\hat{x}^k-\frac{\hbar}{2}\hat{\theta}^k$ \\
      && \\ 
      && \\
 $\hat{\boldsymbol{p}}_0$& $=$ & $\hat{\Omega} + \frac{1}{2c} \hat{E}$ \\
     &&\\
 $\hat{\boldsymbol{p}}_k$&$=$&$\hat{p}_k + \frac{\hbar}{2} \hat{\lambda}_k$   
  \end{tabular}     
  &  
   \begin{tabular}{ccc} 
      $\hat{\boldsymbol{t}}^\prime$& $=$ & $ \hat{t} + \frac{1}{2} \hat{\tau}$    \\
        &   &  \\
    $\hat{\boldsymbol{x}}^{\prime k}$&$=$& $\hat{x}^{k} + \frac{\hbar}{2}\hat{\theta}^k$ \\
      && \\ 
      && \\
 $\hat{\boldsymbol{p}}_0^\prime$& $=$ & $\hat{\Omega} - \frac{1}{2c} \hat{E}$ \\
     &&\\
 $\hat{\boldsymbol{p}}_k^\prime$&$=$&$\hat{p}_k - \frac{\hbar}{2} \hat{\lambda}_k$   
  \end{tabular}   
  \\
   &     &   \\
   \cline{1-3}   
  \end{tabular}
\caption{ Operators in the  \emph{Implicitly Covariant Hilbert Phase} space (ICP) 
where $( \hat{t}, \hat{\tau}, \hat{\Omega} , \hat{E}, \hat{x}^k , \hat{p}_k , \hat{\lambda}_k , \hat{\theta}^k ) $ 
represent the ICP Bopp operators. }
\label{tab-BoppOperators}
\end{table}
obeying  
\begin{align}
\label{commutator-E-tau}
   {[} \hat{t} , \hat{E}  {]} = -i \hbar, &\qquad  {[} \hat{\Omega} , \hat{\tau}  {]} = -i \hbar  ,\\
 {[}  \hat{x}^{j} , \hat{\lambda}_{k} {]} = -i \delta^j_{\,\,\,k}, & \qquad
 {[}  \hat{p}_{j} , \hat{\theta}^{k} {]} = -i \delta^k_{\,\,\,j}, 
\end{align}
where all the other commutators vanish, in particular 
$[\hat{x}^k, \hat{p}_j] = 0$.  A graphical illustration 
of the relation between the time variables  $\boldsymbol{t}-\boldsymbol{t}^\prime$ and $t-\tau$ 
is shown in Fig.~\ref{fig:time-variables}.
Adding and substracting Eqs. 
(\ref{ICHP-vonNeumann-1}) and (\ref{ICHP-vonNeumann-2}), and utilizing the Bopp operators,  
 we obtain the von-Neumann equation in the ICP space  
\begin{align}
\label{ICP2-vonNeumann-1}
\hat{E} |P\rangle \gamma^0 
  &=  \overrightarrow{H} \left( \hat{t} - \frac{\hat{\tau}}{2} , \hat{x}^k-\frac{\hbar}{2}\hat{\theta}^k  ,
   \hat{p}_k + \frac{\hbar}{2} \hat{\lambda}_k  \right) 
 |P\rangle \gamma^0  \\
 &-  |P \rangle  \gamma^{0}  
    \overleftarrow{H} \left( \hat{t} +  \frac{\hat{\tau}}{2}, 
  \hat{x}^k + \frac{\hbar}{2}\hat{\theta}^k
  , \hat{p}_k - \frac{\hbar}{2} \hat{\lambda}_k  \right), \nonumber \\
 \label{ICP2-vonNeumann-2}
 2c \hat{\Omega} |P\rangle \gamma^0 
   &=  \overrightarrow{H} \left( \hat{t} - \frac{\hat{\tau}}{2} , \hat{x}^k-\frac{\hbar}{2}\hat{\theta}^k  ,
   \hat{p}_k + \frac{\hbar}{2} \hat{\lambda}_k  \right) 
 |P\rangle \gamma^0  \\
 &+  |P \rangle  \gamma^{0}  
    \overleftarrow{H} \left( \hat{t} +  \frac{\hat{\tau}}{2}, 
  \hat{x}^k + \frac{\hbar}{2}\hat{\theta}^k
  , \hat{p}_k - \frac{\hbar}{2} \hat{\lambda}_k  \right). \nonumber
\end{align}
$\hat{E}$ and $\hat{\Omega}$  can be realized in terms of differential 
operators as
\begin{align}
\hat{t} &= t        \qquad   \hat{E}   = i \hbar \frac{\partial }{\partial t} ,\\
\hat{\tau} &= \tau  \qquad   \hat{\Omega} = i \hbar  \frac{\partial }{\partial \tau},
\end{align}
turning Eqs. (\ref{ICP2-vonNeumann-1}) and (\ref{ICP2-vonNeumann-2}) into a system of two differential
equations that can be solved by either propagating along $t$  while keeping
$\tau$  fixed, or moving along $\tau$  with $t$ constant. 
In particular, setting $\tau=0$ in Eq. (\ref{ICP2-vonNeumann-1}),  we obtain
 the relativistic von-Neumann equation  in the
\emph{{\bf S}liced {\bf C}ovariant Hilbert {\bf P}hase space} (SCP)
\begin{align}
\label{SCP-vonNeumann}
i \hbar \frac{d}{d t} |P\rangle \gamma^0 
  &=  \overrightarrow{H} \left( \hat{t}  , \hat{x}^k-\frac{\hbar}{2}\hat{\theta}^k  ,
   \hat{p}_k + \frac{\hbar}{2} \hat{\lambda}_k  \right) 
 |P\rangle \gamma^0  \\
 &-  |P \rangle  \gamma^{0}  
    \overleftarrow{H} \left( \hat{t} , 
  \hat{x}^k + \frac{\hbar}{2}\hat{\theta}^k
  , \hat{p}_k - \frac{\hbar}{2} \hat{\lambda}_k  \right). \nonumber
\end{align}
It is well known that a Lorentz
transformation mixes the space and
time degrees of freedom, as recapitulated in Appendix
\ref{Sec:LorentzCovariance}. In particular, the time-evolution of the state in a
different reference frame corresponds to a different slicing 
in the $t-\tau$ plane. Therefore, the state propagated by
Eq. (\ref{SCP-vonNeumann}) with  $\tau=0$ does not
contain enough information to deduce the observations from a
different inertial frame of reference. 
Nevertheless, Eq. (\ref{SCP-vonNeumann}) represents a consistent 
relativistic equation of motion describing  dynamics from the particular 
frame of reference (corresponding to the $\tau=0$ slice) free of any nonphysical artifacts, 
e.g., superluminal propagation.
A schematic illustration of slicing dynamics at $\tau=0$ is shown in Fig. \ref{fig:time-slice}.
Note that equations of motion containing two time
variables also appear in non-relativistic dynamics \cite{manko2012}.

Using Table \ref{tab-BoppOperators}, we rewrite Eq. (\ref{SCP-vonNeumann})  in the Hilbert Spinorial space 
\begin{align}
\label{NCHS-vonNeumann}
  i \hbar \frac{d }{d t} \hat{P}\gamma^0 =&  
  [ H( t  , \hat{\boldsymbol{x}}^k  ,  \hat{\boldsymbol{p}}_k  ) , \hat{P}\gamma^0 ].
\end{align}
Note that this equation resembles Eq. (\ref{eq1}) with $\mathcal{D}=0$. 
In other words, we obtain a straightforward relativistic extension of the density matrix formalism 
for the Dirac equation.
Migdal \cite{Migdal-Bremsstrahlung-PairProduction-1956} employed Eq. (\ref{NCHS-vonNeumann}) 
to describe the effect of multiple scattering on Bremsstrahlung and pair production. 
%In addition Ref. \cite{schreilechner2015single} to develop a finite difference propagator.

\section{Relativistic Wigner function}\label{Sec:RelativisticWigner}

This section is devoted to study specific representations of the   von-Neumann equation  
in the SCP space (\ref{SCP-vonNeumann}) in order to derive the time-evolution 
of the relativistic Wigner function.

\begin{figure}
  \centering
      \includegraphics[width=0.7\hsize]{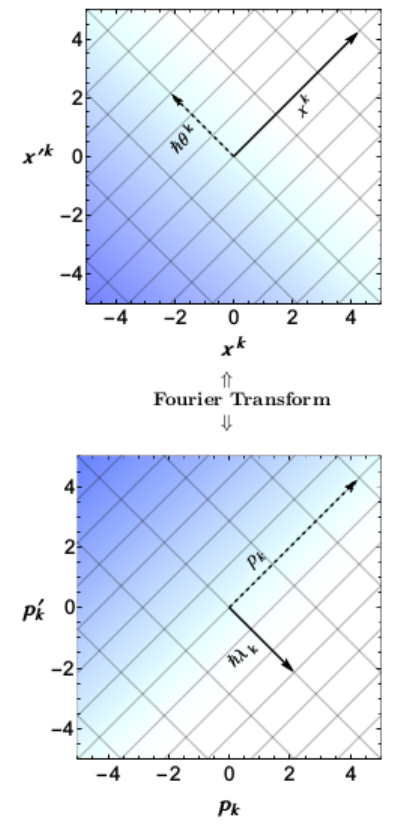}
 \caption{(Color online) Relation between the double configuration ($x^{k}-\theta^{k}$) and the double 
momentum ($\lambda_{k}-p_{k}$) spaces as defined in Table \ref{tab-BoppOperators}. The dashed axes along $p_k$ 
and $\lambda_k$ indicate that they are related via a direct  Fourier transform. The solid axes along $x^k$ 
and $\theta^k$ indicate a similar connection. These relations are also schematically presented 
in  Eq. (\ref{Fourier-transforms}). }
\label{tab:xxpp}    
\end{figure}

%\begin{table}
%
%\begin{tabular}{ccc}
%     & \includegraphics[width=.7\hsize]{positionVars.pdf} & 
%      \\
%     &  \hspace{20pt} $\Uparrow$ & \\
%     & \hspace{38pt}{\bf Fourier Transform} & \\
%     &  \hspace{20pt} $\Downarrow$ & \\     
%      \\
%     & \includegraphics[width=.7\hsize]{momentumVars.pdf} & 
%\end{tabular}
%\caption{(Color online) Relation between the double configuration ($x^{k}-\theta^{k}$) and the double 
%momentum ($\lambda_{k}-p_{k}$) spaces as defined in Table \ref{tab-BoppOperators}. The dashed axes along $p_k$ 
%and $\lambda_k$ indicate that they are related via a direct  Fourier transform. The solid axes along $x^k$ 
%and $\theta^k$ indicate a similar connection. These relations are also schematically presented 
%in  Eq. (\ref{Fourier-transforms}). }
%\label{tab:xxpp}
%\end{table}

Following Table \ref{tab-BoppOperators}, there are four representations of interest:
\begin{itemize}
\item
The \emph{double configuration space} is defined by setting
\begin{align}
  \hat{x}^k = x^k,\quad  \hat{\theta}^k = \theta^k,\quad
  \hat{\lambda}_k = i \frac{\partial }{ \partial x^k},\quad \hat{p}_k = -i \frac{\partial}{ \partial \theta^{k}}.
\end{align}
Hence, the equation of motion (\ref{SCP-vonNeumann}) becomes
\begin{align}
  i \hbar  \frac{\partial B \gamma^0  }{\partial t} =& 
 \overrightarrow{H} \left( t ,  x^k - \frac{\hbar}{2} \theta^k   ,   \hat{p}_k + \frac{\hbar}{2} \hat{\lambda}_k  \right) 
 B \gamma^0  - \nonumber \\
 & B \gamma^0 \overleftarrow{H} \left( t  ,  x^k + \frac{\hbar}{2} \theta^k
    ,  \hat{p}_k - \frac{\hbar}{2} \hat{\lambda}_k   \right), 
\end{align}
where  $B$ is defined as the relativistic Blokhintsev function
\begin{align}
  B\gamma^0 = \frac{1}{\sqrt{ \hbar } } \langle x^k, \theta^k | P \rangle \gamma^0 
   =  \langle  x^k - \frac{\hbar}{2} \theta^k | \hat{P}\gamma^0 |  x^k + \frac{\hbar}{2} \theta^k   \rangle.
\end{align}
For pure states, $B$ is expressed in terms of the four-column Dirac spinor $\psi$ as
\begin{align}
B (t,x^k,\theta^k)\gamma^0 = \psi(t, x^k - \frac{\hbar}{2} \theta^k  ) \psi^{\dagger}(t,   x^k + \frac{\hbar}{2} \theta^k  ).
\end{align}
Therefore, $B$ is a $4\times 4$ complex matrix-valued function of two degrees of freedom $x-\theta$.
 The non-relativistic version of the Blokhintsev function was introduced
in Refs. \cite{Blokhintsev1940a,Blokhintsev1940,Blokhintsev1941}. 

\item
The \emph{phase space}  is defined by 
\begin{align}
  \hat{x}^k = x^k,\quad  \hat{p}_k = p_k,\quad  \hat{\lambda}_k = i \frac{\partial }{ \partial x^k}, \quad
  \hat{\theta}^k = i \frac{\partial }{ \partial p_k}.
\end{align}
The underlying equation of motion  (\ref{SCP-vonNeumann}) reads 
\begin{align}
  i \hbar  \frac{\partial W \gamma^0  }{\partial t} =& 
 \overrightarrow{H} \left( t ,  x^k - \frac{\hbar}{2} \hat{\theta}^k   ,   p_k + \frac{\hbar}{2} \hat{\lambda}_k  \right) 
 W \gamma^0  - \nonumber \\
 & W \gamma^0 \overleftarrow{H} \left( t ,  x^k + \frac{\hbar}{2} \hat{\theta}^k
    ,  p_k - \frac{\hbar}{2} \hat{\lambda}_k   \right), 
\end{align}
where $W$ is the sought after relativistic Wigner function
\begin{align}
 \label{def-W}
  W\gamma^0 = \frac{1}{2\pi \hbar} \langle x^k, p^k | P \rangle\gamma^0,
\end{align}
which can be recovered from the Blokhintsev function through a Fourier transform
\begin{align}
 \label{def2-W}
 	W(t,x^k,p^k) = \frac{1}{(2\pi)^3} \int B(t,x^k,\theta^k)  \exp( i p \cdot \theta) d^3\theta.
\end{align}
Note that only contravariant components are used in Eqs. (\ref{def-W}) and (\ref{def2-W}). 

\item The \emph{reciprocal phase space} is defined as
\begin{align}
  \hat{x}^k = -i \frac{\partial }{ \partial \lambda_k} ,\quad
  \hat{p}_k = -i \frac{\partial }{ \partial \theta^k} , \quad 
  \hat{\lambda}_k = \lambda_k, \quad
  \hat{\theta}^k = \theta^k.
\end{align}
The corresponding equation of motion is 
\begin{align}
  i \hbar  \frac{\partial \mathcal{A} \gamma^0  }{\partial t} =& 
 \overrightarrow{H}\left( t ,  \hat{x}^k - \frac{\hbar}{2} \theta^k  ,   \hat{p}_k + \frac{\hbar}{2} \lambda_k  \right) 
 \mathcal{A} \gamma^0  - \nonumber \\
 & \mathcal{A} \gamma^0 \overleftarrow{H} \left( t ,  \hat{x}^k + \frac{\hbar}{2} \theta^k
    ,  \hat{p}_k - \frac{\hbar}{2} \lambda_k   \right), 
\end{align}
where $\mathcal{A}$ is the relativistic ambiguity function
\begin{align}
  \mathcal{A}\gamma^0 = \frac{1}{\sqrt{ \hbar } } \langle \lambda^k, \theta^k | P \rangle\gamma^0,
\end{align}
which is recovered from the Blokhintsev function according to
\begin{align}\label{Deff_A}
 	A(t, \lambda^k,\theta^k) =  \int B(t,x^k,\theta^k)  \exp( -i x \cdot \lambda ) d^3x.
\end{align}
\item
The \emph{double momentum space} is introduced as
\begin{align}
   \hat{x}^k = -i \frac{\partial }{ \partial \lambda_k}  ,\quad  \hat{p}_k = p_k,\quad  \hat{\lambda}_k = \lambda, \quad
  \hat{\theta}^k = i \frac{\partial }{ \partial p_k}.
\end{align}
The corresponding equation of motion is 
\begin{align}
  i \hbar  \frac{\partial Z \gamma^0  }{\partial t} =& 
 \overrightarrow{H} \left( t ,  \hat{x}^k - \frac{\hbar}{2} \hat{\theta}^k  ,   p_k + \frac{\hbar}{2} \lambda_k  \right) 
 Z \gamma^0  - \nonumber \\
 & Z \gamma^0 \overleftarrow{H} \left( t ,  \hat{x}^k + \frac{\hbar}{2} \hat{\theta}^k
    ,  p_k - \frac{\hbar}{2} \lambda_k   \right), 
\end{align}
where 
\begin{align}
  Z\gamma^0 = \frac{1}{\sqrt{ \hbar } }\langle \lambda^k, p^k | P \rangle \gamma^0 
   =  \langle  p^k + \frac{\hbar}{2} \lambda^k | \hat{P}\gamma^0 |  p^k - \frac{\hbar}{2} \lambda^k   \rangle,
\end{align}
which is related with the Wigner function via 
\begin{align}\label{Deff_Z}
 	W(t,x^k,p^k) = \frac{1}{(2\pi)^3} \int Z(t,\lambda^k,p^k)  \exp( i x \cdot \lambda ) d^3\lambda.
\end{align}
Similarly, we also have 
\begin{align}\label{Deff_Z2}
 	\mathcal{A}(t, \lambda^k,\theta^k) =  \int Z(t,\lambda^k, p^k)  \exp( -i p \cdot \theta ) d^3p.
\end{align}
\end{itemize}

In summary, all these four functions are connected through Fourier transforms 
as visualized in the following diagram:
{\large
\begin{align}
  \label{Fourier-transforms}
    \xymatrix{
      W(x,p)  \ar[r]^{ \mathcal{F}_{ x \rightarrow \lambda  } }  &   Z(\lambda,p)   \\
      B(x,\theta)  \ar[u]^{\mathcal{F}_{  \theta \rightarrow  p   }}  \ar[r]^{ \mathcal{F}_{ x \rightarrow \lambda  } }  & 
   \mathcal{A}(\lambda,\theta)  \ar[u]_{ \mathcal{F}_{  \theta \rightarrow  p   }}     }
\end{align}} 
where vertical arrows denote the direct ${\mathcal{F}_{  \theta \rightarrow  p     }} $ Fourier transforms
while horizontal arrows indicate the direct  $ \mathcal{F}_{ x \rightarrow \lambda  } $ Fourier transforms.
A similar diagram can be drawn in terms of the inverse Fourier transforms 
as
{\large
\begin{align}
  \label{Fourier-transforms-inverse}
    \xymatrix{
      W(x,p)  \ar[d]^{\mathcal{F}_{  p \rightarrow  \theta   }}   &  
       Z(\lambda,p)  \ar[l]_{ \mathcal{F}^{ \lambda \rightarrow x  } }  \ar[d]^{\mathcal{F}_{  p \rightarrow  \theta   }}  \\
      B(x,\theta)   & 
   \mathcal{A}(\lambda,\theta)    \ar[l]_{ \mathcal{F}^{ \lambda \rightarrow x  } }   }
\end{align}} 

Since the relativistic Wigner function $W$ is a $4 \times 4$ complex matrix,
its visualization is cumbersome. Nevertheless, most of
the information is contained in \cite{campos2014violation}
\begin{align}
	W^0(t,x^k, p^k) &\equiv {\rm Tr}\, [W(t,x^k , p^k)\gamma^0]/4. \label{W-0}
\end{align}
In fact, this zero-th component is sufficient to obtain the probability density $j^0 \equiv \psi^{\dagger}(t,x^k) \psi(t,x^k)  $ as
\begin{align}
	\int  W^0(t,x^k, p^k) d^3p &= \psi^{\dagger}(t,x^k) \psi(t,x^k)   \label{x_marginal_W} \\
	\int  W^0(t,x^k, p^k) d^3x &= \widetilde{\psi}^{\dagger}(t,p^k) \widetilde{\psi}(t,p^k), \label{p_marginal_W} 
\end{align}
where $\widetilde{\psi}$ is the Dirac spinor in the momentum representation, i.e. the Fourier transform of $\psi$.

Equations (\ref{x_marginal_W}) and (\ref{p_marginal_W}) reveal that the zero-th component of the relativistic 
Wigner function (\ref{W-0}) acts as a quasi-probability distribution -- a real valued non-positive function, 
whose marginals coincide with the coordinate and momentum probability densities, respectively.

\section{Open system interactions}\label{Sec:OpenSysInter}

Inspired by non-relativistic quantum mechanics [see Eq. (\ref{eq1})],
we add a dissipator to the relativistic von Neumann equation  (\ref{NCHS-vonNeumann})
to account for open system dynamics
 \begin{align}
\label{non-cov-open-system}
i\hbar \frac{d }{d t} \hat{P}\gamma^0  = {[} 
 H(t, \hat{\boldsymbol{x}}^k ,  \hat{\boldsymbol{p}}^k   ) , \hat{P}\gamma^0  {]}
 + i \hbar \mathcal{D}( \hat{P}\gamma^0 ).   
\end{align}
We note that Eq.  (\ref{non-cov-open-system}) does not need 
to comply with relativistic covariance. Nevertheless, this is not a deficiency when dealing
with environments such as thermal baths that are already furnished with a preferred frame of reference. 

Motivated by the treatment of quantum dephasing in non-relativity \cite{gardiner2004quantum},
we propose to include the following dissipator in Eq. (\ref{non-cov-open-system})  
\begin{align}
\label{dephasing-operator}
\mathcal{D}[\hat{P} \gamma^0 ] &= - \frac{D}{\hbar^2} [\hat{\boldsymbol{x}}^{k} 
                                   , [ \hat{\boldsymbol{x}}^k , \hat{P}\gamma^0 ] ] , 
\end{align}
where $D$ is the decoherence coefficient controlling the dephasing intensity and no summation on $k$ is
implied. In non-relativistic systems this interaction is utilized to
describe the loss of coherence due to the interaction with an
environment associated with a thermal bath
\cite{Caldeira1983587,PhysRevLett.80.4361,PhysRevLett.88.040402,zurek1991decoherence}. In addition,
a system undergoing continuous measurements in position follows the same dynamics
\cite{1402-4896-1998-T76-027,PhysRevA.60.2700}. A model similar to Eq. (\ref{dephasing-operator}) describes the effect of multiple scattering 
on Bremsstrahlung and pair production in high energy limit of the incident electron \cite{Migdal-Bremsstrahlung-PairProduction-1956}.

The dynamical effect of an interaction
can be characterized by calculating the time derivative of the expectation
value of an observable $\hat{O}$
\begin{align}
 \frac{d}{dt} \langle \hat{O} \rangle = 
Tr\left[ \frac{d }{d t}( \hat{P}\gamma^0 ) \hat{O}   \right].
\end{align}
Assuming that the equation of motion is of the form
\begin{align}
 \frac{d}{dt} \hat{P} \gamma^0 = \mathcal{M}( \hat{P}\gamma^0  ),
\end{align}
the time derivative of $\langle \hat{O} \rangle$ is expressed as follows
\begin{align}
\label{dOdt}
 \frac{d}{dt} \langle \hat{O} \rangle =  
Tr\left[ \mathcal{M}( \hat{P}\gamma^0 ) \hat{O}   \right] 
=  Tr\left[ \hat{P}\gamma^0  \mathcal{M}^{\dagger}(\hat{O})   \right],
\end{align}
where $\mathcal{M}^{\dagger}$ is the adjoint operator  of $\mathcal{M}$ with respect to the Hilbert-Schmidt scalar product.

The particular dephasing dissipator (\ref{dephasing-operator}) is self-adjoint,
\begin{align}
\mathcal{D}^{\dagger}[ \hat{O} ] &=\mathcal{D}[ \hat{O} ];
\end{align}
as a result, 
\begin{align}
 \label{Dephasing-first-order}
 \mathcal{D}^{\dagger}[ \hat{\boldsymbol{x}}^k ] = \mathcal{D}^{\dagger}[ \hat{\boldsymbol{p}}^k ]  = 0. 
\end{align}
This means that the dephasing does not change the Heisenberg equations of motion for position and momentum observables. 
The open system interaction affects the dynamics of the second order momentum  
\begin{align}
  \mathcal{D}^{\dagger}[ \hat{\boldsymbol{x}}^k\hat{\boldsymbol{x}}^j  ]=0 \quad
 \mathcal{D}^{\dagger}[ \hat{\boldsymbol{p}}^k \hat{\boldsymbol{p}}^j ] = 2 D \delta^{kj},\quad
  \mathcal{D}^{\dagger}[ \hat{\boldsymbol{x}}^k\hat{\boldsymbol{p}}^j  ]=0,
\end{align}
which in turn leads to a momentum wavepacket broadening.  Moreover,
considering that the free Dirac Hamiltonian (\ref{Dirac-Hamiltonian}) is
linear in momentum, we obtain from Eqs. (\ref{Dephasing-first-order})  and (\ref{dOdt}) 
\begin{align}
  \frac{d}{dt} \left \langle \gamma^0\gamma^k \boldsymbol{\hat{p}}_k + m c \gamma^0  \right \rangle = 0.
\end{align}
In other words, the energy is conserved under the action of the
dephasing dissipator (\ref{dephasing-operator}). This is in stark
contrast to non-relativistic dephasing, which is characterized by
monotonically increasing energy.

The classical limit of dephasing  (\ref{dephasing-operator})  is diffusion.
Relativistic extensions of diffusion face fundamental challenges
\cite{Dunkela2009}. For instance, large values of $D$ may induce 
dynamics leading to superluminal propagation, which breaks down
the causality of the Dirac equation (see, e.g., Theorem 1.2 of
Ref. \cite{thaller1992dirac}). The length-scale of diffusion is
$\sqrt{\langle x^2 \rangle} = \sqrt{2Dt}$; hence, the characteristic
speed $\sqrt{\langle x^2 \rangle}/t = \sqrt{2D/t} $ must be smaller
than the speed of light. The shortest time interval for which the
single particle picture is valid $t \sim \hbar / ( 2mc^2 )$, i.e., the \emph{zitterbewegung} time scale. Considering all these
arguments, we obtain the constrain: $D \ll \hbar / (4m)$, or
equivalently, $4D/c \ll \lambdabar$ (where $\lambdabar = \hbar / (mc)$
is the reduced Compton wavelength) in order to maintain causal dephasing dynamics.

This dephasing interaction (\ref{dephasing-operator}) can be expressed in the  SCP space, leading to a very simple 
expression \cite{PhysRevA.92.042122}
\begin{align}
\label{dephasing-operator-Hilbert-phase-space}
 \frac{\partial }{\partial t}   
 \langle x^j \theta^j | P \rangle = - D\, \theta^k \theta^k \delta^{kj}  \langle x^j \theta^j | P \rangle,
\end{align}
which is convenient for numerical propagation, as shown in Sec. \ref{Sec:NumAlg}.

\section{Numerical algorithm}\label{Sec:NumAlg}
Stimulated by the resurgent interest in the Dirac equation, a
plethora of propagation methods were recently developed
\cite{mocken2008fft,bauke2013computational,Fillion-Gourdeau2011,Hammer201440,hammer2014single,beerwerth2015krylov}.
However, to the best our knowledge Ref. \cite{schreilechner2015single} is the only work 
devoted to propagation of the relativistic von Neumann equation (\ref{NCHS-vonNeumann}), albeit without 
open system interactions.
The purpose of this section is to develop an effective numerical
algorithm to propagate the master equation (\ref{non-cov-open-system}) describing
quantum dephasing (\ref{dephasing-operator}). 
The computational effort with the proposed algorithm scales as the square of the Dirac equation propagation complexity.
This algorithmic development enables the relativistic Wigner function simulations, which were previously hindered by
the complexity of the underlying  integro-differential equations \cite{vasak1987quantum,hakim1978covariant}.

The evolution governed by  Eq. (\ref{NCHS-vonNeumann})
\begin{align}
  i \hbar \frac{d }{d t} \hat{Q} =&  
  [ H( t  , \boldsymbol{x}^k  ,  \hat{\boldsymbol{p}}_k  ) , \hat{Q} ],
\end{align}
with $ \hat{Q} = \hat{ P} \gamma^0$ is equivalent to  
\begin{align}
 \hat{Q}_{t+dt} = e^{-i dt H(t, \hat{\boldsymbol{x}}, \hat{\boldsymbol{p}} )/\hbar} \hat{Q}_t 
 e^{ i dt H(t, \hat{\boldsymbol{x}}, \hat{\boldsymbol{p}} )/\hbar},
\end{align}
where $dt$ is an infinitesimal time step.

Considering that the Hamiltonian can be decomposed as
\begin{align}
\hat{H} &= K( \hat{\boldsymbol{p}} ) +  V(\hat{ \boldsymbol{x} }), \\ 
  K(\hat{\boldsymbol{p}}) &=   c \alpha^k \hat{\boldsymbol{p}}^k   + mc^2 \gamma^0/2,  \\
  V(\hat{\boldsymbol{x}}) &=    e A^0(t, \hat{\boldsymbol{x}}^k) -  e \alpha^k  A^k(t, \hat{\boldsymbol{x}}^k)  + mc^2 \gamma^0/2,
\end{align}
where the mass term contributes to both $K(\hat{\boldsymbol{p}})$ and 
$ V(\hat{\boldsymbol{x}})$. The first order splitting with error $O(dt^2)$ is then
\begin{align}
 \hat{Q}_{t+dt} =  e^{-i dt V( \hat{\boldsymbol{x}} )/\hbar}  e^{-i dt K( \hat{\boldsymbol{p}} )/\hbar} \hat{Q}_t 
 e^{ i dt K( \hat{\boldsymbol{p}} )/\hbar} e^{ i dt V( \hat{\boldsymbol{x}} )/\hbar},
\end{align}
which implies a two step propagation 
\begin{align}
 \hat{Q}^{1/2} &=  e^{-i dt K( \hat{\boldsymbol{p}} )/\hbar} \hat{Q}_t 
 e^{ i dt K( \hat{\boldsymbol{p}} )/\hbar} \\
\hat{Q}_{t+dt} &= e^{-i dt V( \hat{\boldsymbol{x}} )/\hbar}   \hat{Q}^{1/2}  e^{ i dt V( \hat{\boldsymbol{x}} )/\hbar}
\end{align}

Using Eqs.  (\ref{rule1}) and (\ref{rule2}) we move to SCP  
\begin{align}
 |Q^{1/2} \rangle &= e^{-i dt \overrightarrow{K}( \hat{\boldsymbol{p}} )/\hbar } |Q_t \rangle 
 e^{ i dt \overleftarrow{K}( \hat{\boldsymbol{p}}^{\prime} )/\hbar }, \\
 |Q_{t+dt} \rangle &= e^{-i dt \overrightarrow{V}( \hat{\boldsymbol{x}} )/\hbar } |Q^{1/2} \rangle 
 e^{ i dt \overleftarrow{V}( \hat{\boldsymbol{x}}^{\prime} )/\hbar}.
\end{align}
Note that $| Q_t \rangle$ is a complex $ 4\times 4$ matrix reflecting the spinor degrees of freedom. 
The arrows can be eliminated by choosing suitable bases   
\begin{align}
 \langle \boldsymbol{p} \boldsymbol{p}^{\prime} |Q^{1/2} \rangle &
 = e^{-i dt K( \hat{\boldsymbol{p}} )/\hbar }  \langle \boldsymbol{p} \boldsymbol{p}^{\prime}|Q_t \rangle 
 e^{ i dt K( \hat{\boldsymbol{p}}^{\prime} )/\hbar }, \\
    \langle \boldsymbol{x} \boldsymbol{x}^{\prime} |Q^{1/2} \rangle &=
   \mathcal{F}_{ \boldsymbol{p} \boldsymbol{p}^{\prime}  \rightarrow  \boldsymbol{x} \boldsymbol{x}^{\prime}    } 
    \langle \boldsymbol{p} \boldsymbol{p}^{\prime} |Q^{1/2} \rangle,  \\
 \langle \boldsymbol{x} \boldsymbol{x}^{\prime} |Q_{t+dt} \rangle &= e^{-i dt V( \hat{\boldsymbol{x}} )/\hbar }
 \langle \boldsymbol{x} \boldsymbol{x}^{\prime} |Q^{1/2} \rangle 
 e^{ i dt V( \hat{\boldsymbol{x}}^{\prime} )/\hbar  }, \\
 \langle \boldsymbol{p} \boldsymbol{p}^{\prime} |Q_{t+dt} \rangle &=
   \mathcal{F}^{ \boldsymbol{x} \boldsymbol{x}^{\prime}  \rightarrow  \boldsymbol{p} \boldsymbol{p}^{\prime}    } 
    \langle \boldsymbol{x} \boldsymbol{x}^{\prime} |Q_{t+dt} \rangle, 
\end{align}
where  $ \mathcal{F}_{ \boldsymbol{p}
  \boldsymbol{p}^{\prime} \rightarrow \boldsymbol{x}
  \boldsymbol{x}^{\prime} } $ and $ \mathcal{F}^{ \boldsymbol{x}
  \boldsymbol{x}^{\prime} \rightarrow \boldsymbol{p}
  \boldsymbol{p}^{\prime} } $ stand for Fourier transforms from the
momentum representation to the position representation and
vice versa. Considering that the state is a $4\times 4$ matrix, the
Fourier transform is independently applied to each matrix component. 
From the computational perspective, the fast Fourier transform is employed. 
Further details about the phase space propagation via the fast Fourier transform can be found
in Sec. III of Ref. \cite{PhysRevA.92.042122}.

Having described the propagation algorithm in SCP $( \hat{\boldsymbol{x}}^k, \hat{\boldsymbol{x}}^{k\prime} ,
\hat{\boldsymbol{p}}^k, \hat{\boldsymbol{p}}^{k\prime} )$, one can apply a similar strategy to the Bopp operators
 $(\hat{x}^k, \hat{p}^k, \hat{\theta}^k , \hat{\lambda}^k)$ (see Table \ref{tab-BoppOperators}). There are multiple advantages of the latter representation. 
Importantly, some open system interactions (e.g., the dephasing model explained 
in detail in Sec. \ref{Sec:OpenSysInter}) take simpler forms in terms of $(\hat{x}^k, \hat{p}^k, \hat{\theta}^k ,
\hat{\lambda}^k)$. 
The momentum and coordinate grids in  $( \hat{\boldsymbol{x}}^k,
\hat{\boldsymbol{x}}^{k\prime} , \hat{\boldsymbol{p}}^k,
\hat{\boldsymbol{p}}^{k\prime} )$ are interdependent
such that if the discretization step size $\boldsymbol{dx}$ and the grid amplitude of $\boldsymbol{x}$ are specified, then
the momentum increment $\boldsymbol{dp}$ and the amplitude of $\boldsymbol{p}$ are fixed and vice versa. 
However, the momentum and position grids in  $(\hat{x}^k, \hat{p}^k, \hat{\theta}^k ,
\hat{\lambda}^k)$ are independent, thus allowing the flexibility to choose
$dx$, $dp$, and amplitudes of $x$ and $p$, in order to resolve the quantum dynamics of interest.

The following equation of motion is obtained from Eq. (\ref{SCP-vonNeumann}):
\begin{align}
i \hbar \frac{d}{d t} |Q\rangle  
  &=  \overrightarrow{K} \left( \hat{p}_k + \frac{\hbar}{2} \hat{\lambda}_k  \right) 
 |Q\rangle
 -  |Q \rangle    
    \overleftarrow{K} \left( \hat{p}_k - \frac{\hbar}{2} \hat{\lambda}_k  \right), \nonumber \\
 &+  \overrightarrow{V} \left( \hat{x}_k + \frac{\hbar}{2} \hat{\theta}_k  \right) 
 |Q\rangle 
 -  |Q \rangle    
    \overleftarrow{V} \left( \hat{x}_k - \frac{\hbar}{2} \hat{\theta}_k  \right). 
\end{align}
The first order splitting leads to the two step propagation
\begin{align}
 | Q^{1/2} \rangle &=  e^{-\frac{i dt}{\hbar} \overrightarrow{K}( \hat{p} + \frac{\hbar}{2}\hat{\lambda} ) } | Q_t \rangle 
 e^{ \frac{i dt}{\hbar} \overleftarrow{K}( \hat{p} - \frac{\hbar}{2}\hat{\lambda} )  }, \\
| Q_{t+dt} \rangle &= e^{-\frac{i dt}{\hbar} \overrightarrow{V}( \hat{x} - \frac{\hbar}{2} \hat{\theta}  ) }   | Q^{1/2} \rangle 
  e^{\frac{i dt}{\hbar} \overleftarrow{V}( \hat{x} + \frac{\hbar}{2} \hat{\theta}  ) }.
\end{align}
\begin{figure}
  \centering
      \includegraphics[width=0.9\hsize]{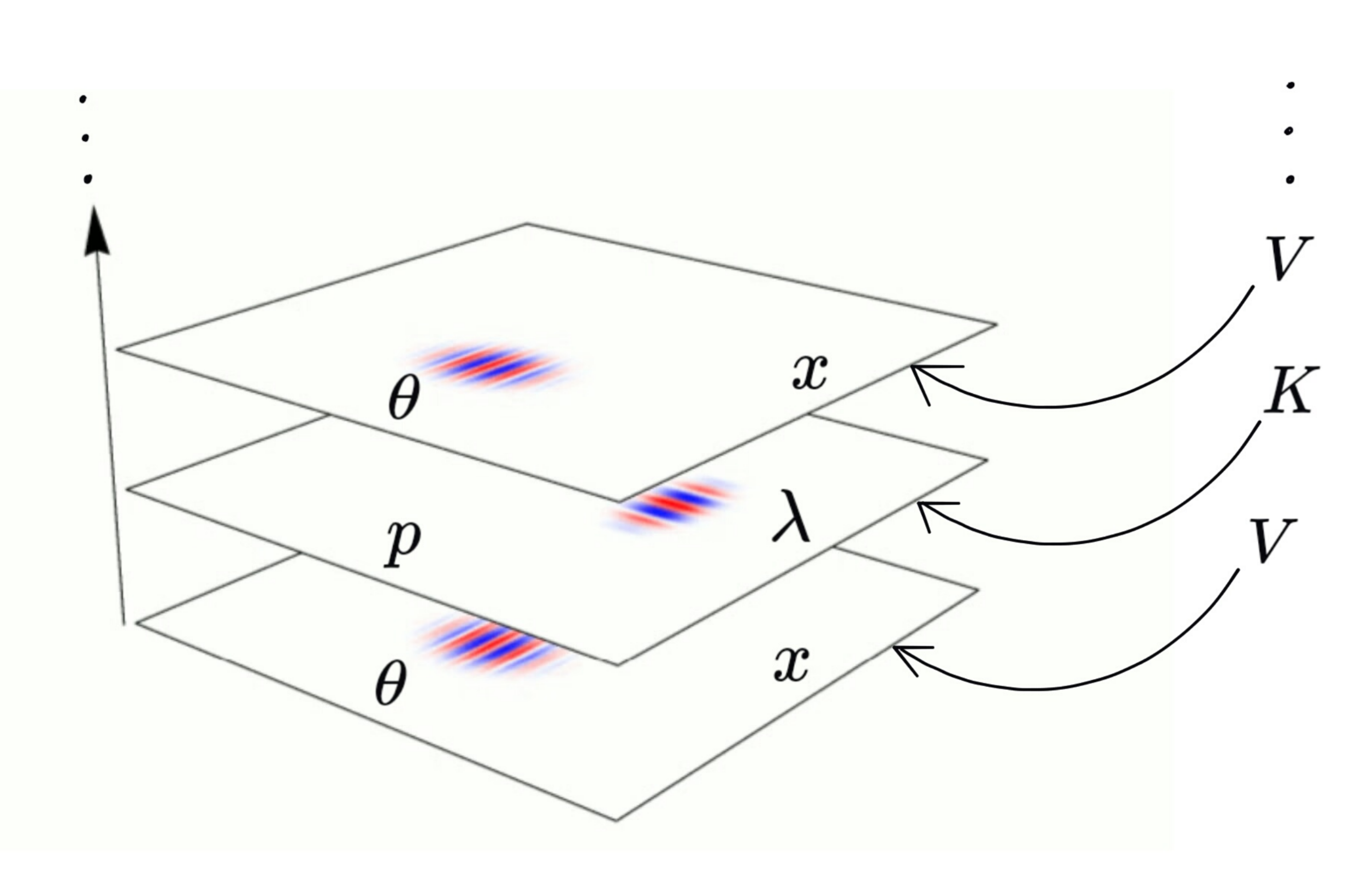}
      \caption{ (Color online) Schematic representation of the iterative steps to propagate the quantum state 
according to Eqs. (\ref{Eq:SplitOp1})-(\ref{Eq:SplitOp4}).  }
  \label{fig:SplitOperatorScheme}
\end{figure}
The employment of the appropriate basis at each step removes the need for arrows 
\begin{align}
\label{Eq:SplitOp1}
\langle \lambda p  | Q^{1/2} \rangle &=  e^{-\frac{i dt}{\hbar} K( p + \frac{\hbar}{2}\lambda ) } \langle \lambda p | Q_t \rangle 
 e^{ \frac{i dt}{\hbar} K( p- \frac{\hbar}{2}\lambda )  }, \\
  \langle x \theta  | Q^{1/2} \rangle &=  \mathcal{F}^{  \lambda p \rightarrow   x \theta   }  \langle \lambda p  | Q^{1/2} \rangle, \\
\langle x \theta | Q_{t+dt} \rangle &= e^{- \frac{i dt}{\hbar} V( x - \frac{\hbar}{2} \theta  ) } \langle x \theta  | Q^{1/2} \rangle 
  e^{\frac{i dt}{\hbar} V( x+ \frac{\hbar}{2} \theta  )  }, \label{Eq:step3}\\
  \langle \lambda p  | Q^{1/2} \rangle &=  \mathcal{F}_{   x \theta \rightarrow \lambda p  }  \langle \lambda p  | Q^{1/2} \rangle,
\label{Eq:SplitOp4}
\end{align}
where the Fourier transform conform with Eq.(\ref{Fourier-transforms}) and Eq. (\ref{Fourier-transforms-inverse}) 
according to   
 \begin{align}
 \mathcal{F}_{ x\theta \rightarrow \lambda p  } &\equiv 
 \mathcal{F}_{x\rightarrow \lambda} \mathcal{F}_{\theta \rightarrow p} 
= \mathcal{F}_{\theta \rightarrow p} \mathcal{F}_{x\rightarrow \lambda}, \\
 \mathcal{F}^{ \lambda p  \rightarrow  x \theta  } &\equiv \mathcal{F}^{ \lambda  \rightarrow  x  } \mathcal{F}^{p  \rightarrow  \theta  } =  \mathcal{F}^{p  \rightarrow  \theta  } \mathcal{F}^{ \lambda  \rightarrow  x  }  .
\end{align} 
A schematic view of the sequence of steps (\ref{Eq:SplitOp1})-(\ref{Eq:SplitOp4})  is shown in Fig. \ref{fig:SplitOperatorScheme}.
Note that to maintain consistency, the propagator must be  solely expressed in terms of 
contravariant components, e.g., 
\begin{align} 
  K( p \pm \frac{\hbar}{2}\lambda  ) & =    c \alpha^k \left( p^k \pm   \frac{\hbar}{2}\lambda^k \right )   + mc^2 \gamma^0/2.  
\end{align}
The matrix exponentials in Eq. (\ref{Eq:SplitOp1}) can be evaluated analytically.
For instance, assuming a two dimensional quantum system (ignoring  $x^3$ and $p^3$)
we obtain
\begin{align}
 e^{  -\frac{i dt}{\hbar} \left[ c \alpha^k p^k  + mc^2 \gamma^0 \right]  } = 
    \begin{pmatrix} \mathcal{K}_{11} & 0 & 0 &  \mathcal{K}_{14}       \\
                        0 &  \mathcal{K}_{11} &  \mathcal{K}_{23}  & 0  \\
                        0 &  \mathcal{K}_{32} &  \mathcal{K}_{11}^*  & 0  \\
                      \mathcal{K}_{41}& 0     & 0      &  \mathcal{K}_{11}^*
    \end{pmatrix}, 
\label{expK-matrix}
\end{align}
with
\begin{align}
  \mathcal{K}_{11} &=   \cos( c dt F/\hbar ) - i m c \frac{ \sin( c dt F/\hbar )}{F}, \\
  \mathcal{K}_{14} &=   \frac{\sin( c dt F/\hbar )}{F} \left( -i p^1 -  p^2    \right),   \\
  \mathcal{K}_{23} &=   -U_{14}^*,  \\
  \mathcal{K}_{32} &=   U_{14},  \\
  \mathcal{K}_{41} &=   -U_{14}^*, \\
  F &= \sqrt{ (mc)^2 + (p^1 )^2 + (p^2 )^2  }.
\end{align}

Likewise, the exponential in Eq. (\ref{Eq:step3}) yields
\begin{align}
 e^{  -\frac{i dt}{\hbar} \left[ \alpha^{\mu} e A_{\mu}  + mc^2 \gamma^0 \right]  } = 
  e^{-\frac{i e A^0 dt}{\hbar} }  \begin{pmatrix} \mathcal{A}_{11} & 0 &  \mathcal{A}_{13} &  \mathcal{A}_{14}       \\
                        0 &  \mathcal{A}_{11} &  \mathcal{A}_{23}  & \mathcal{A}_{24}   \\
                       \mathcal{A}_{31} &  \mathcal{A}_{32} &  \mathcal{A}_{11}^*  & 0  \\
                      \mathcal{A}_{41}&  \mathcal{A}_{42}     & 0      &  \mathcal{A}_{11}^*
    \end{pmatrix}, 
\label{expA-matrix}
\end{align}
with 
\begin{align}
  \mathcal{A}_{11} &= \cos(dt G/\hbar) -i m c^2 \frac{\sin( dt G/\hbar)  }{ G }, \\
  \mathcal{A}_{31} &=  \mathcal{A}_{13} = i A^3  \frac{\sin( dt G/\hbar)  }{ G },   \\
  \mathcal{A}_{41} &=  \mathcal{A}_{23}  = (-A^2 + i A^1)  \frac{\sin( dt G/\hbar)  }{ G },\\
  \mathcal{A}_{32} &=   \mathcal{A}_{14}  = - \mathcal{A}_{41}^*, \\
  \mathcal{A}_{42} & = \mathcal{A}_{24}  =  \mathcal{A}_{31}^*, \\
   G &=  \sqrt{ (mc^2)^2 + (A^1 )^2 + (A^2 )^2 + (A^3)^2 }.
\end{align}

Having described the propagation for closed system Dirac evolution, 
we  now proceed to introduce quantum dephasing (\ref{dephasing-operator}), a particular open system interaction. 
According to Eq. (\ref{dephasing-operator-Hilbert-phase-space}), the dephasing dynamics enters into the
exponential of the potential energy, thereby modifying  the propagation step (\ref{Eq:step3})  as
\begin{align}
 \label{Eq:step3-dephasing}
\langle x \theta | Q_{t+dt} \rangle = 
 e^{- \frac{i dt}{\hbar} \tilde{V}( x - \frac{\hbar}{2} \theta  ) } \langle x \theta  | Q^{1/2} \rangle 
  e^{ \frac{i dt}{\hbar} \tilde{V}( x+ \frac{\hbar}{2} \theta  ) },
\end{align}
with
\begin{align}
 - \frac{i dt}{\hbar} \tilde{V}\left( x \pm \frac{\hbar}{2} \theta  \right) &=
  -\frac{i dt}{\hbar} V \left( x \pm \frac{\hbar}{2} \theta  \right) - \frac{D dt}{2} \theta^2. 
\end{align}
The replacement of Eq. (\ref{Eq:step3}) by Eq. (\ref{Eq:step3-dephasing}) is mathematically equivalent to
Gaussian filtering along the momentum axis (i.e., convolution with a Gaussian in momentum) 
of the coherently propagated $W(t,x^1,p^1)$. This simple interpretation of the dephasing dynamics plays a crucial role in Sec. \ref{Sec:MajoranaIlustrations}. 

The presented  algorithm can be implemented with the 
resources of a typical desktop computer and are well suited for GPU computing \cite{Klockner2012157}. 
In particular, the illustration in the next section were executed 
with a Nvidia graphics card Tesla C2070.

\section{Majorana Spinors}\label{Sec:MajoranaIlustrations}

Hereafter, assuming a one dimensional dynamics, the Wigner function takes the functional form 
 $W(t,x^1,p^1)$. Furthermore, natural units ($c=\hbar=1$) are used throughout. In this section 
we employ a $512 \times 512$  grid for $x^1$ and $p^1$ as well as a time step $dt=0.01$.  Animations 
of simulations can be found in Ref. \cite{SupplementalM}.

Majorana spinors, characterized for being their own antiparticles, 
are the subject of interest in a broad range of fields including high energy physics,
quantum information theory and solid state physics \cite{wilczek2009majorana}. 
In particular, the solid state counterpart of the relativistic Majorana spinors 
is known to be robust against perturbations and imperfections 
due to peculiar topological features  \cite{albrecht2016exponential}. 
  
\begin{figure}
  \centering
      \includegraphics[width=1.\hsize]{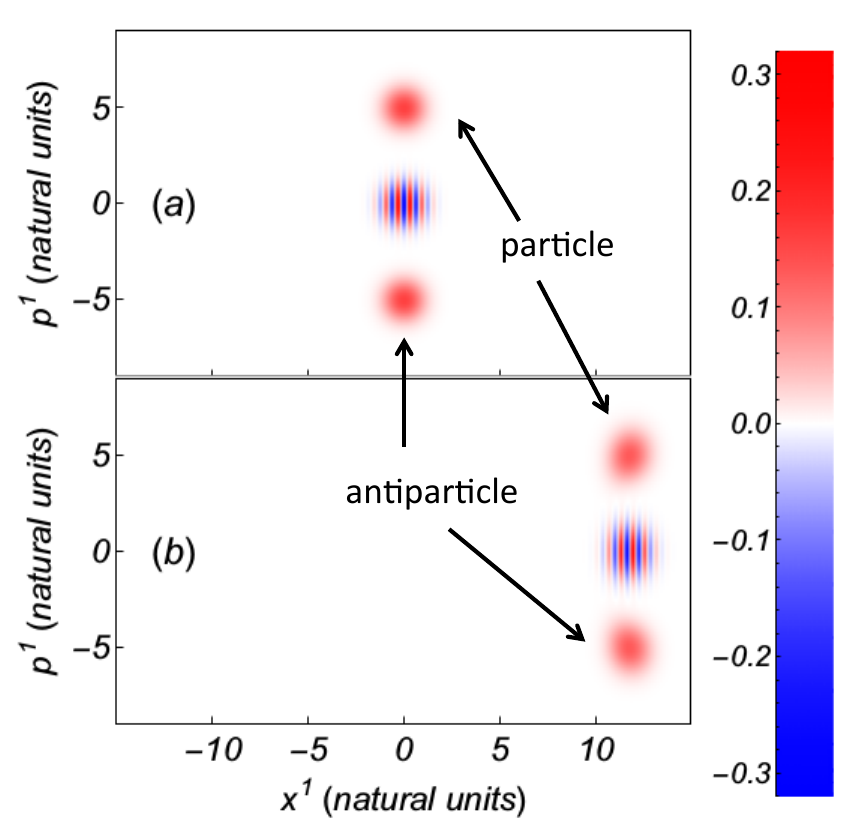}
      \caption{(Color online) The relativistic Wigner function
        $W^0(t,x^1,p^1)$ for a Majorana spinor $\psi^M_{+}$ associated
        with the spinor in Eq. (\ref{psi_initial}) at (a) $t=0$ and (b)
        at $t = 12$. Note that the particle undergoes dephasing with coefficient $D=0.01$, without 
        an external electromagnetic field. An animated illustration can be found in \cite{SupplementalM}.  }
  \label{fig:Majorana}
\end{figure}

In this section we study the dynamics of the original Majorana spinor \cite{majorana1937teoria} 
in the presence of  dephasing noise (\ref{dephasing-operator}). Let  
\begin{align}
 \psi = \begin{pmatrix} \psi_1 \\ \psi_2 \\ \psi_3 \\ \psi_4
        \end{pmatrix}
\end{align}
be an arbitrary spinor, then there are two underlying Majorana states 
(see, e.g.,  Chapter 12, page 165 of Ref. \cite{lounesto2001clifford})
\begin{align}
\label{MajoranaFormula}
 \psi^{M}_{ \pm } = \begin{pmatrix} \psi_1 \\ \psi_2 \\ \psi_3 \\ \psi_4
        \end{pmatrix} \pm 
        \begin{pmatrix} -\psi_4^* \\ \psi_3^* \\ \psi_2^* \\ -\psi_1^*
        \end{pmatrix}.
\end{align}
In particular, we propagate the Majorana spinor $\psi^{M}_{+}$ [shown in Fig.~\ref{fig:Majorana}(a)] 
obtained from 
\begin{align}
\label{psi_initial}
   \psi_0 =  e^{ -\frac{(x^1)^2}{2} + i x^1 \tilde{p}^1 } 
   ( \tilde{p}^0 + m c,0,0, \tilde{p}^1 )^T,
\end{align}
with  $\tilde{p}^0 = \sqrt{ (\tilde{p}^1)^2 + (m c)^2  }$ and the numerical values  $\tilde{p}^1 = 5$, $m=1$ and the dephasing coefficient $D=0.01$ in natural units. 
The resulting  time propagation of  $\psi^{M}_{+}$  is shown in  Fig. \ref{fig:Majorana} (b).

Figure \ref{fig:Majorana} reveals that the particle-antiparticle
superposition of the Majorana state generates a strong interference in
the phase space, which survives an even very intense dephasing
interaction. The reason of such robustness is that both the particle
component (with a positive momentum) and the antiparticle component
(with a negative momentum) move in \emph{parallel} along the positive
spatial direction.
 This is in agreement with the interpretation 
of antiparticles as particles moving backwards in time. The interference fringes, consisting
of negative and positive stripes, also remain parallel to the momentum
axis. Considering the remark after Eq. (\ref{Eq:step3-dephasing}), the
action of dephasing is equivalent to the Gaussian filtering along the
$p^1$ axis only. This mixes negative values with negative, positive values with positive, 
but never positive with negative values of the Wigner function.
Hence, this leaves the interference stripes invariant. 
In other words, free Majorana spinors evolve in a decoherence-free subspace \cite{lidar1998decoherence}.

\begin{figure}
  \centering
      \includegraphics[width=1.\hsize]{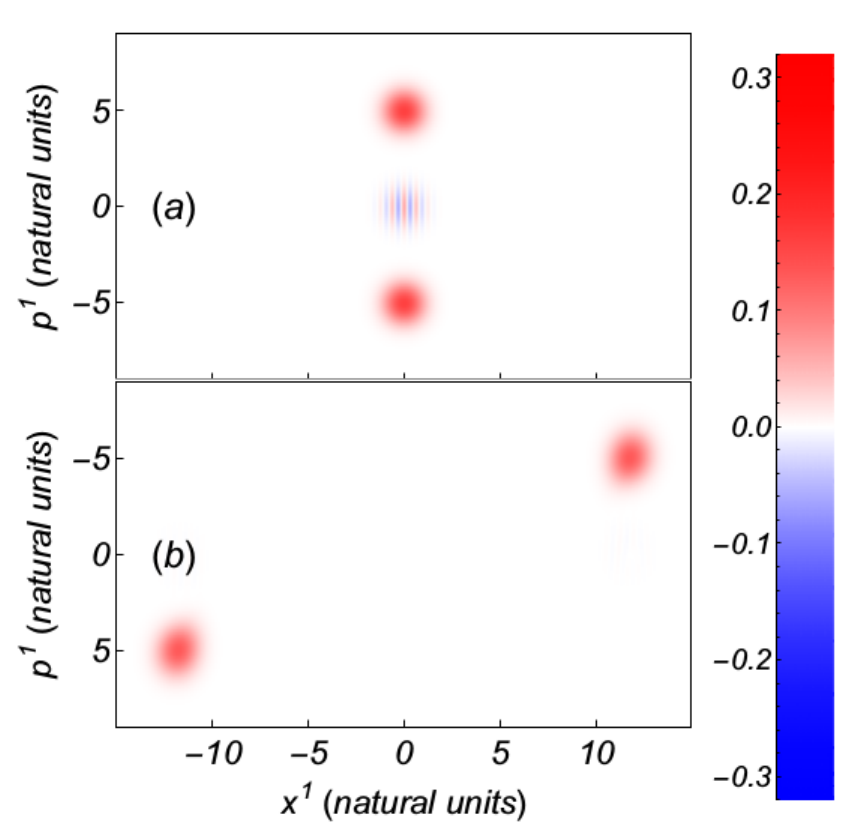}
      \caption{(Color online)  The relativistic Wigner function
        $W^0(t,x^1,p^1)$ for a particle-particle superposition
        corresponding to the spinor in
        Eq. (\ref{psi_initial_schrodinger_cat}) at (a) $t=0$ and (b) 
         at $t = 12$. Note that the particle undergoes dephasing with coefficient $D=0.01$, without 
        an external electromagnetic field. An animated illustration can be found in \cite{SupplementalM}. }
  \label{fig:CatStateV}
\end{figure}

The described Majorana state dynamics is fundamentally different from the evolution
of a cat-state, i.e., a particle-particle superposition.  For example,
up to a normalization factor, consider the following initial
cat-state, composed of mostly  particles: 
\begin{align}
\label{psi_initial_schrodinger_cat}
	\psi_0 =   e^{  -\frac{(x^1)^2}{2}  } \left[e^{ i x^1 \tilde{p}^1  }   + e^{  -i x^1 \tilde{p}^1 } \right] 
	( \tilde{p}^0 + m c , 0 , 0, \tilde{p}^1 )^T.
\end{align}
Figure \ref{fig:CatStateV} depicts the evolution of this state under the influence 
of the same dephasing interaction as in  Fig. \ref{fig:Majorana}. Contrary to the Majorana case,
the  negative momentum components of the cat state are made of particles; therefore, 
we observe  in Fig. \ref{fig:CatStateV} that they move along the negative spatial
direction. The interference stripes connecting the positive (moving to the right) and negative (moving to the left) 
momentum components no longer remain parallel with respect to the $p^1$ axis. 
Thus, dephasing occurs as the Gaussian filtering averages over positive and negative stripes,
thereby washing interferences out.     

We note that the distortion from the original Gaussian character 
 of particle and atiparticle states at initial time  in Figs. \ref{fig:Majorana}
and \ref{fig:CatStateV} is due to the momentum dispersion.

The total integrated negativity of the Wigner function
\begin{align}
\label{Negativity-formula}
 N(t) = \int_{ W^0(t, x^1, p^1)  <0} W^0(t, x^1, p^1) dx^1 dp^1
\end{align}
is widely  regarded as a measurement of the 
quantum coherence because interferences are associated 
with  Wigner function's negative values.
 \begin{figure}
  \centering
      \includegraphics[width=1.\hsize]{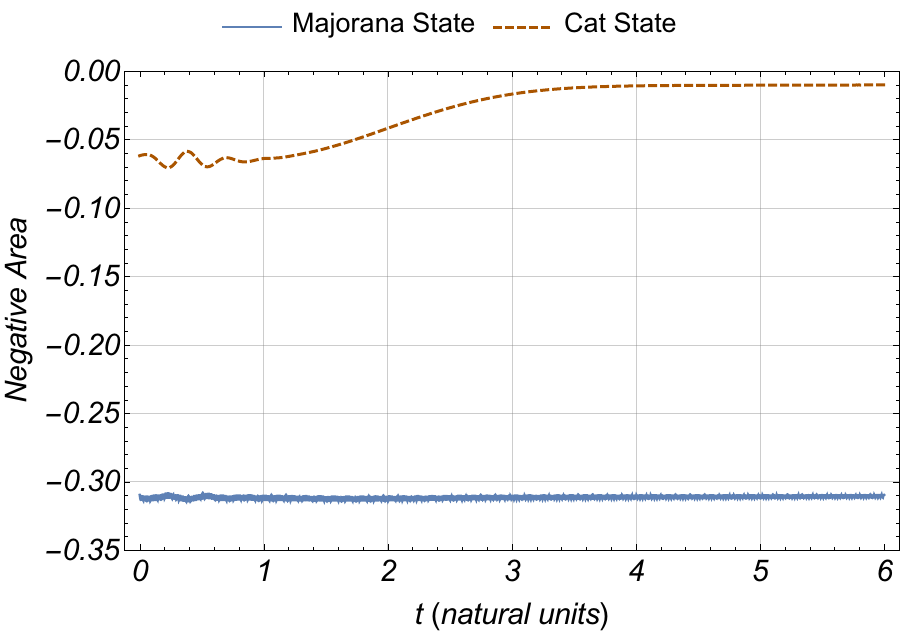}
      \caption{(Color online) Negativity  (\ref{Negativity-formula}) of the Majorana state
        (a particle-antiparticle superposition) in solid line
        corresponding to the free evolution presented in
        Fig. \ref{fig:Majorana} in comparison with the negativity of
        the cat state (a particle-particle superposition) in dashed lines
        corresponding to Fig. \ref{fig:CatStateV}. }
  \label{fig:Negativity}
\end{figure}
Figure \ref{fig:Negativity} shows that the negativity
of the cat state reduces, while the negativity of the Majorana state
is constant. Moreover, the negativity of the free
Majorana spinor remains constant even for extreme values of the
decoherences. Therefore, this robustness is not a perturbative
effect with respect to the dephasing coefficient $D$. 
Note that Majorana spinor's initial negativity is more pronounced than that 
of the cat state (Fig. \ref{fig:Negativity}). Hence, Majorana states 
are more coherent than cat-states.

\begin{figure}
  \centering
      \includegraphics[width=1.\hsize]{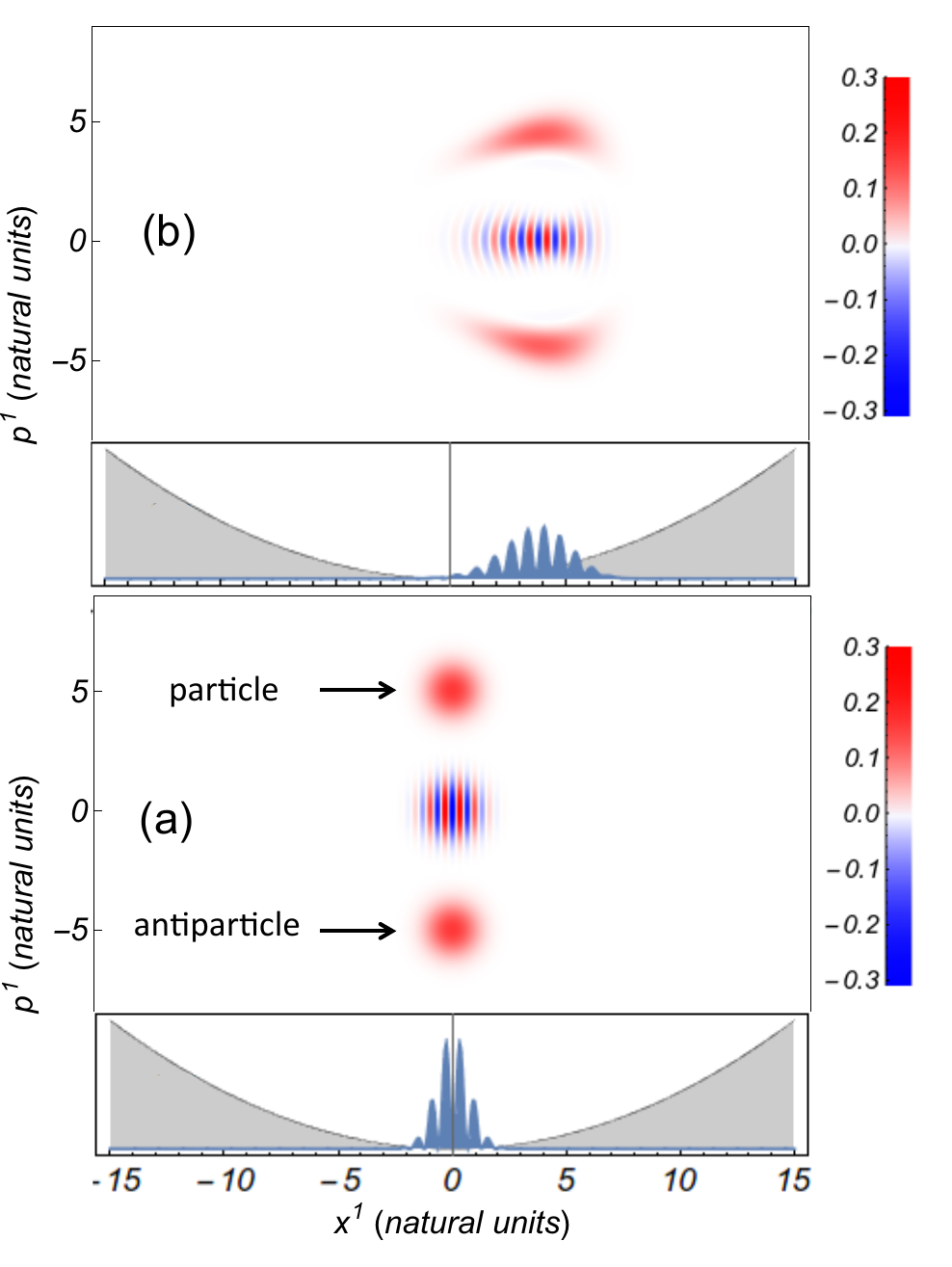}
      \caption{(Color online) (a) Initial Majorana state extracted from (\ref{psi_initial}),
along with its marginal distribution in position where the gray area represents the underlying mass modulated potential
$m \rightarrow m + 0.05 (x^1)^2 $.
 (b) Propagated Majorana state at time $t=14.$ An animated illustration can be found in \cite{SupplementalM}. }
  \label{fig:MassMajorana}
\end{figure}

\begin{figure}
  \centering
      \includegraphics[width=1.\hsize]{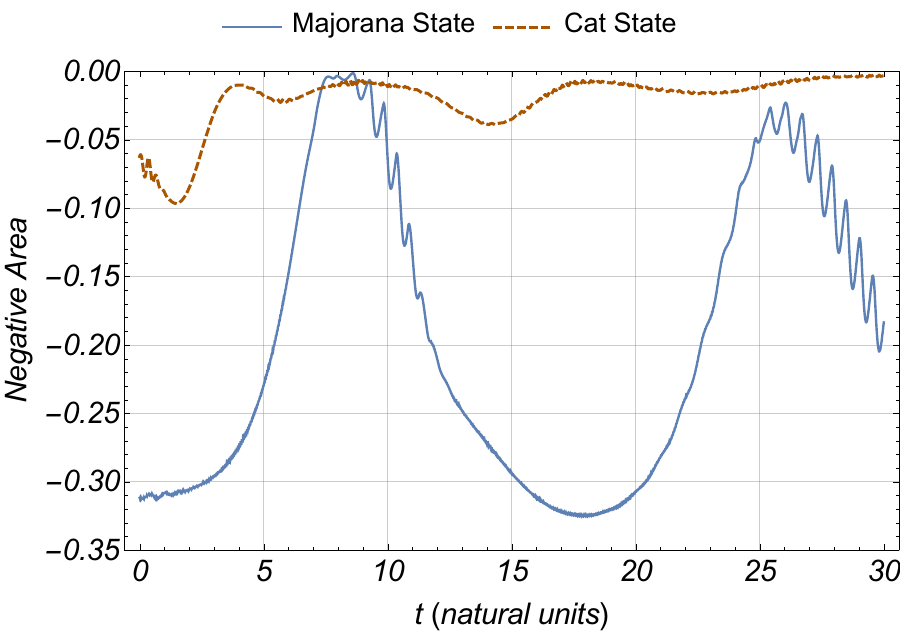}
      \caption{(Color online) Negativity of the Majorana state of Fig. \ref{fig:MassMajorana} in solid line, 
        compared to the  negativity of the corresponding cat state.  }
  \label{fig:NegativityMassMajorana}
\end{figure}

\begin{figure}
  \centering
      \includegraphics[width=0.9\hsize]{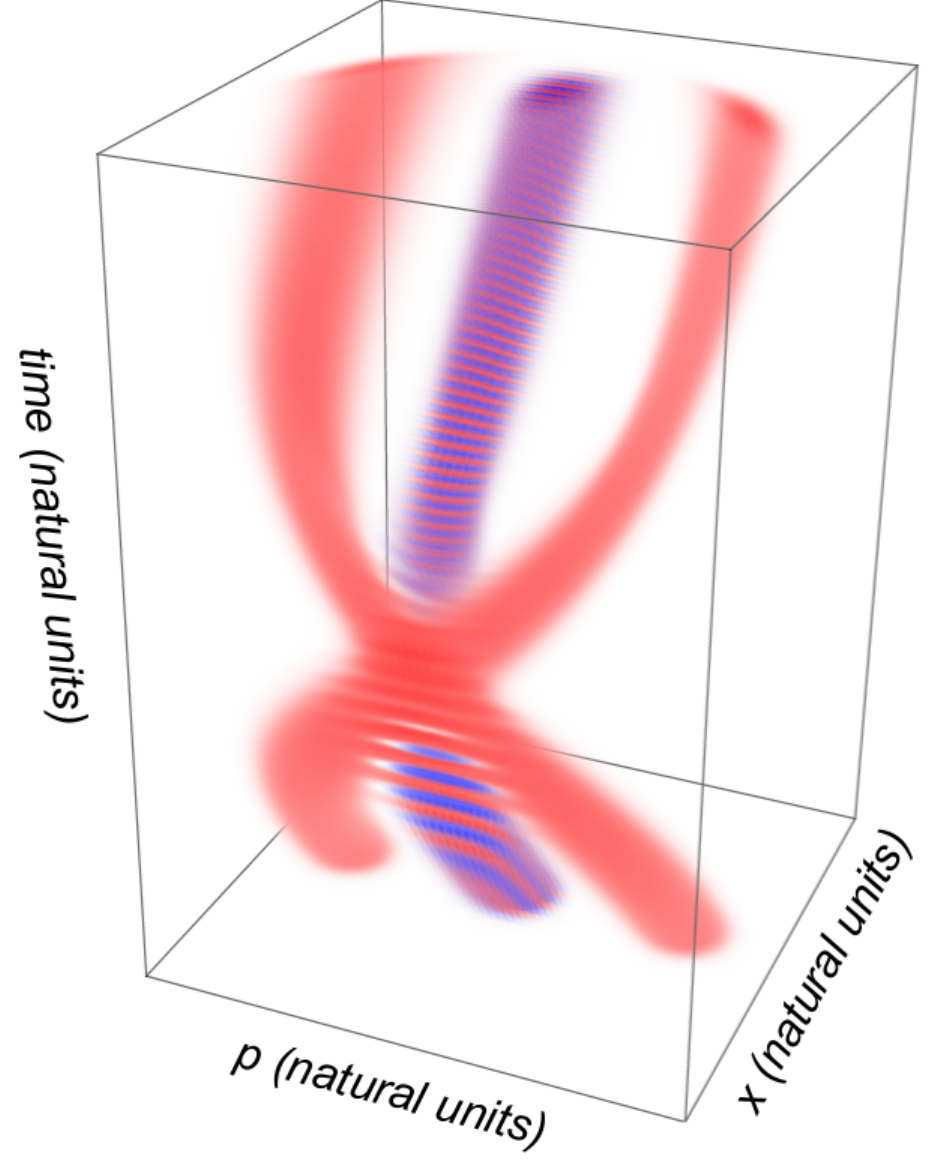}
      \caption{(Color online) Time stacked relativistic Wigner function  ($ 0 \leq t \leq 20 $) for 
the Majorana dynamics shown in Fig. \ref{fig:MassMajorana}. 
The interferences, located in the middle, remain robust all along the evolution  despite 
of the presence of significant quantum decoherence. The inteferences contain regions of negative value in blue. 
The integrated negativity (\ref{Negativity-formula})  as a function in time 
is shown in Fig. \ref{fig:NegativityMassMajorana}. }
  \label{fig:MassMajorana3D}
\end{figure}

\begin{figure}
  \centering
      \includegraphics[width=0.9\hsize]{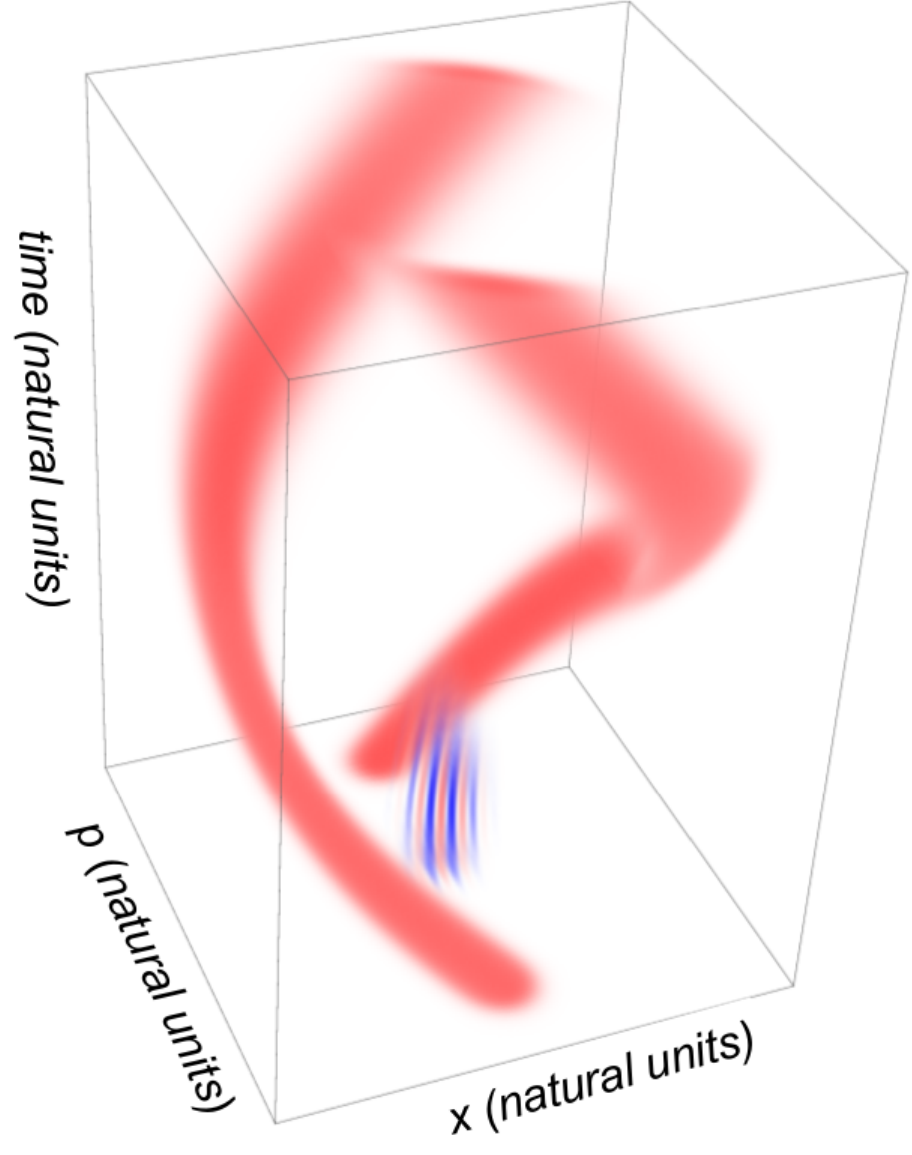}
      \caption{(Color online) Time stacked relativistic Wigner function ($ 0 \leq t \leq 20 $)  for a cat state
        evolving in the same potential as the Majorana spinor in Fig. \ref{fig:MassMajorana3D}. The
        interferences, fade shortly after the initiation of the propagation due to the action of quantum
        decoherence. The integrated negativity (\ref{Negativity-formula}) as a function in time is shown in
        Fig. \ref{fig:NegativityMassMajorana}. }
  \label{fig:CatState3D}
\end{figure}

Having studied free evolution, we now proceed to a Majorana
state evolving under the influence of the spatially modulated 
mass   $m
\rightarrow m + 0.05 (x^1)^2 $. This type of system also maintains 
a high coherence despite significant dephasing  $D=0.01$.  The initial Majorana
state is shown in Fig. \ref{fig:MassMajorana} (a) while the propagated state
at time $t=14.$ is shown in Fig. \ref{fig:MassMajorana} (b). 
The latter figure shows that interference is preserved. 
A comparison of the negativities for Majorana and cat-states as functions of time 
are shown in Fig. \ref{fig:NegativityMassMajorana}, where 
the Majorana state negativity oscillates albeit with some decay, 
which is much slower than the cat-state decay. 
Figure \ref{fig:MassMajorana3D}, showing the full Wigner dynamics, sheds light
on the revival of the Majorana's negativity: When the particle and antiparticle 
components merge and separate, the negativity disappears and appears, respectively.  
Furthermore, Majorana's dynamics seems to be approximately constrained to a surface 
in the phase space and time, in contrast to the cat-state dynamics 
shown in Fig.  \ref{fig:CatState3D}.     

\section{ Klein Tunneling}\label{Sec:KleinIllustrations}
\begin{figure}
  \centering
      \includegraphics[width=1.\hsize]{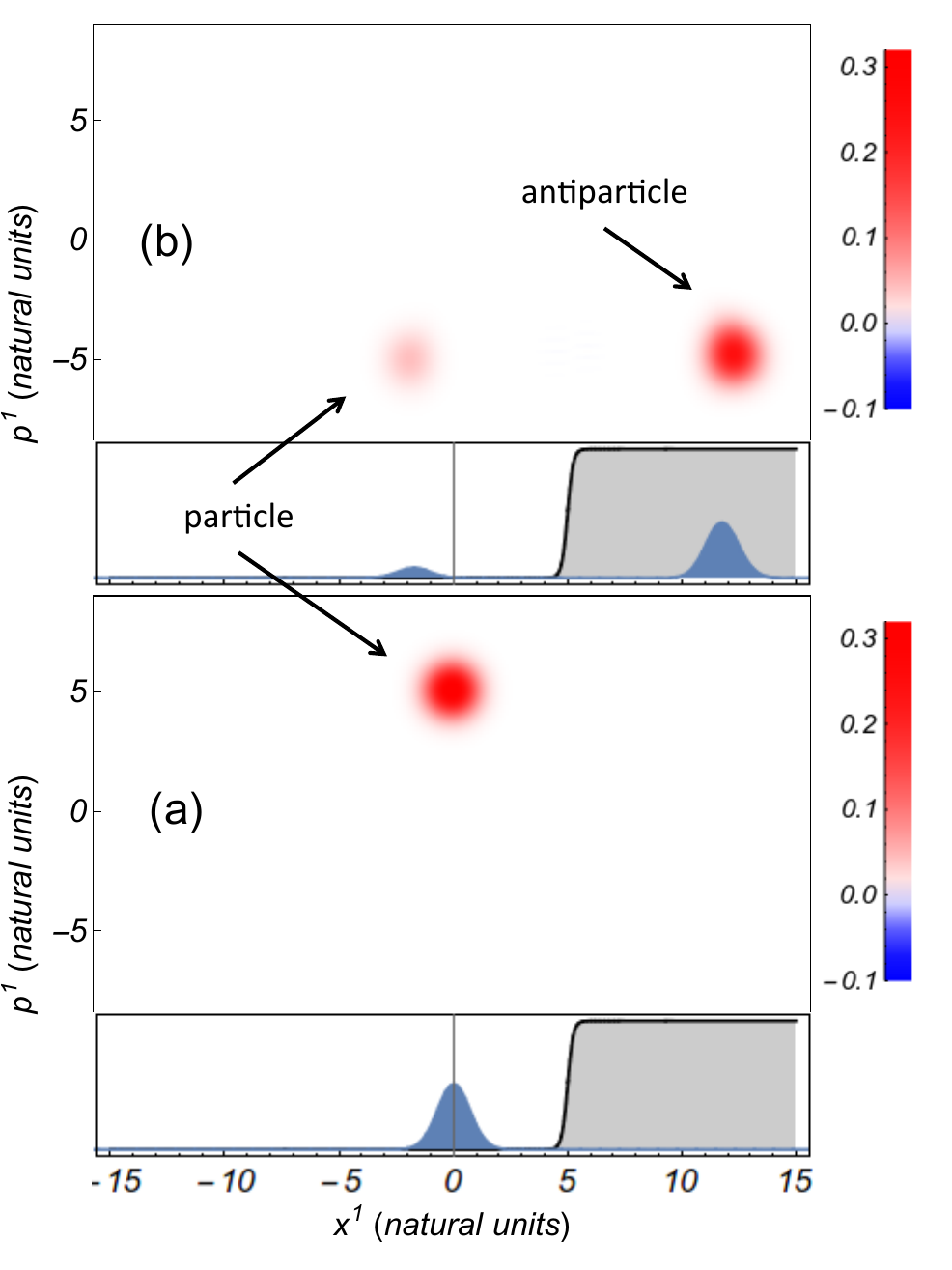}
      \caption{(Color online) Illustration of the Klein paradox in
        terms of the relativistic Wigner function. The step potential 
$A_0 = 10(1 + \tanh[4(x-5)] )/2$.
is depicted as a gray area. The height of the step potential is $V_0=10$ 
        while the energy of the initial wavepacket is $E=5.01$. 
        (a) The initial state $W^0(t=0,x^1,p^1)$ from Eq. (\ref{psi_initial}) with
        $\tilde{p}^1 = 5.$ aimed towards the barrier. (b) Final state of the relativistic Wigner
        function at $t=12$ made of mostly of a negative energy wavepacket (antiparticle) being
        transmitted through the barrier. }
  \label{fig:Klein}
\end{figure}

As the second numerical example, we examine the Klein paradox \cite{greiner2000relativisticIB}, an
unexpected consequence of the Dirac equation, predicting that a positive energy particle colliding with a sharp
potential barrier of the height $V>mc^2$ is transmitted as a negative energy state.  
For example, the initial state  (\ref{psi_initial}) with $\tilde{p}^1 = 5$, $m=1$  is 
shown  in Fig.   \ref{fig:Klein} (a) along with the potential $A_0 = 10(1 + \tanh[4(x-5)]  )/2$.
We observe in Fig. \ref{fig:Klein} (b) that most of the wavepacket has been transmitted as antiparticles. 

\begin{figure}
  \centering
      \includegraphics[width=1.\hsize]{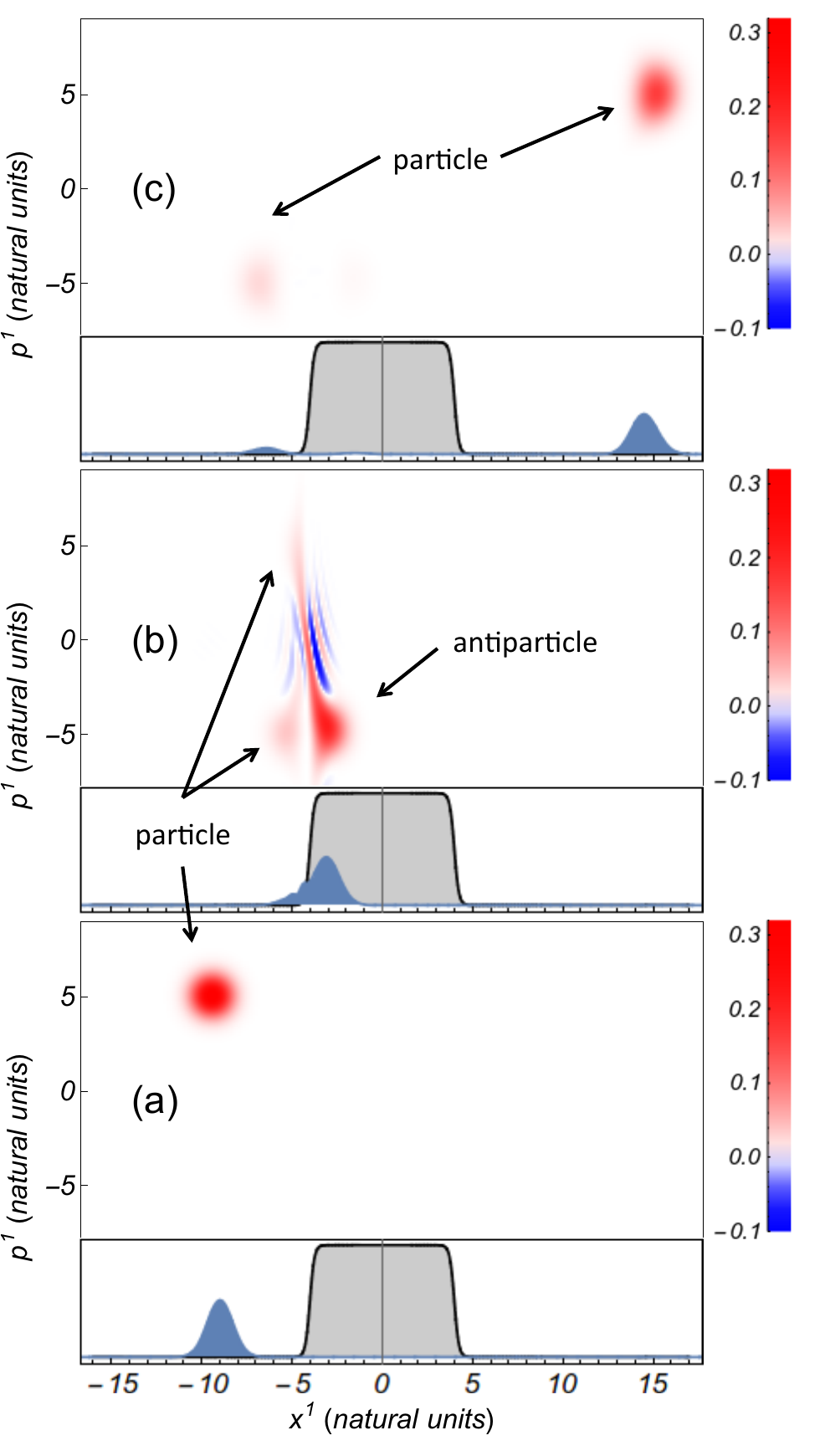}
      \caption{(Color online) Illustration of the Klein tunneling in
        terms of the relativistic Wigner function with the potential
        barrier $A_0 = 5(\tanh[4(x+4)] + \tanh[4(-x+4)] )$ depicted as
        a gray area.  (a) The relativistic Wigner function
        $W^0(t=0,x^1,p^1)$ for the initial state  in Eq. (\ref{psi_initial}) with $\tilde{p}^1 = 5.$ and positioned
        around $x^1=-10$.   (b) The
        relativistic Wigner function at $t=6$ in the process of
        entering the potential and transforming into a negative energy wavepacket (antiparticle).
        (c) The final relativistic Wigner
        function at $t=24$, where most of the initial wavepacket has
        been transmitted as a positive energy wavepacket (particle).  }
  \label{fig:KleinTunneling}
\end{figure}

An important extension of the Klein paradox is the Klein tunneling,
where the step potential is replaced by a finite width barrier. 
In this case, the theoretical prediction specifies a high transmission even for a wide barrier. 
Condensed matter analogies of this phenomenon are a subject of active research \cite{katsnelson2006chiral,young2009quantum}. 
Three snapshots of the Klein tunneling dynamics are shown in Fig. \ref{fig:KleinTunneling},
where (a) corresponds to the positive energy initial state, (b) the state penetrating the
potential barrier as antiparticle, and (c) the final state emerging from the barrier as particle. 

The Dirac particle has a spinorial as well as a configurational degree of freedom. 
The Klein tunneling can be viewed as an interband transition between positive and negative 
energy states \cite{allain2011klein}. Analogous effects exist in non-relativistic dynamics.
In particular, compared to the structureless case,  non-relativitic systems with many degrees of freedom
 manifest many unique peculiarities such as, e.g., transmission 
rate  enhancement \cite{zakhariev1964intensified,bondar2010enhancement} and  directional 
symmetry breaking \cite{Amirkhanov1966}. Thus, the energy exchange between different 
degrees of freedom underlies the counterintuitive dynamics of both the Klein and the 
non-relativistic tunneling of particles with internal structure. 

Furthermore, the Klein tunneling can be interpreted as the Landau-Zener transition between 
positive and negative energy states. This conclusion is obtained, e.g., by comparing 
Eqs. (\ref{diagonalDiracG}) and (\ref{classical_a})  (setting $A^\mu=0$) with Eqs. (19)-(21) 
in Ref. \cite{kane1960zener}. This observation underscores an analogy between solid state and relativistic physics.  
 
\begin{figure}
  \centering
      \includegraphics[width=1.\hsize]{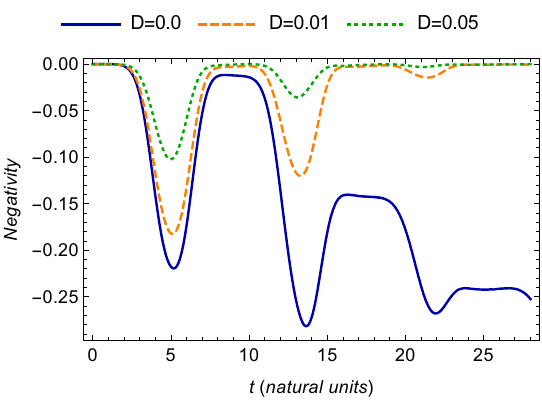}
      \caption{The integrated negative area in Eq. (\ref{Negativity-formula})  as a function of time for
        the Klein tunneling process.  Three different values of the decoherence coefficient are considered for the
        same initial state depicted in   Fig. \ref{fig:KleinTunneling} (a).  The first dip corresponds
        to the first contact of the wave packet with the barrier as
        shown in Fig. \ref{fig:KleinTunneling} (b). The second dip
        corresponds to the main wavepacket emerging from the
        barrier. Other smaller dips appear as a contribution of the
        smaller reflected wavepackets. }
  \label{fig:NegativityKT}
\end{figure}

\begin{figure}
  \centering
      \includegraphics[width=1.\hsize]{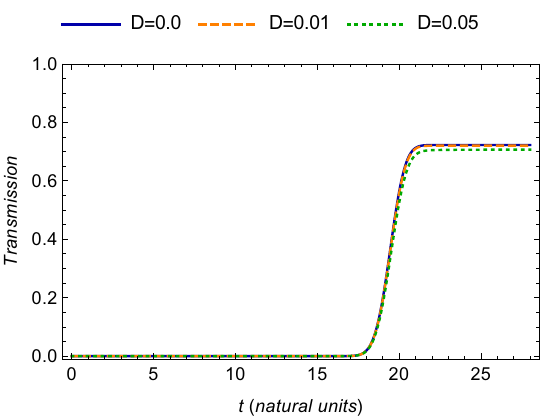}
  \caption{(Color online) The Klein transmission across the potential barrier as a function of time 
for the initial wavepacket shown 
in Fig. \ref{fig:KleinTunneling} (a), indicating a weak dependence on the dephasing intensity. }
  \label{fig:transmissionKT}
\end{figure}

\begin{figure}
  \centering
      \includegraphics[width=1.\hsize]{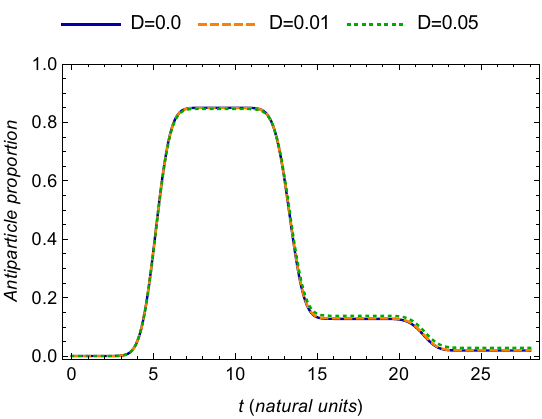}
      \caption{(Color online) 
        The antiparticle proportion as a function of time for three different
        values of the decoherence coefficient in the Klein tunneling process. 
        The initial state composed of mostly particles 
        is shown in Fig.~\ref{fig:KleinTunneling} (a).
        The first high plateau  corresponds to the period of time when most of the wavepacket travels
        within the potential barrier as an antiparticle.  }
  \label{fig:ParticleKT}
\end{figure}

Simulations with different values of the dephasing coefficient $D$  have been performed
in order to investigate the effect of decoherence  on the final transmission. Figure \ref{fig:NegativityKT} 
depicts the integrated negativity (\ref{Negativity-formula}) as a function of time for
three different values of $D$. The evolution
without decoherence generates high negativity that indicates 
interference between the larger transmitted and smaller reflected wavepackets. 
In the same figure we observe that the decoherence eliminates negativity 
at later stages of the propagation.
Nevertheless, the effect of decoherence on the final transmission rate is
small in Fig. \ref{fig:transmissionKT}, where the transmission as a function of time nearly coincides 
for different values of $D$.
We also note a weak dependence of the antiparticle generation on the dephasing coefficient 
as shown in Fig. \ref{fig:ParticleKT}.
Contrary to non-relativistic quantum dynamics \cite{zurek1991decoherence,1402-4896-1998-T76-027,RevModPhys.75.715,
  PhysRevLett.80.4361,PhysRevLett.88.040402,PhysRevLett.96.010403,PhysRevA.67.042103,PhysRevA.92.042122},
decoherence in the relativistic regime does not recover a single particle classical description.
Furthermore, we show in Appendix \ref{Sec:ClassLimitDirac}
that \emph{the  limit $\hbar \to 0$ of the Dirac equation 
leads to two classical Hamiltonians}: One describing
particles with a forward advancing clock (i.e., particles), while
the other -- a particle with backward flowing
proper time (i.e., antiparticles). (This limit of the Dirac equation represents an example 
of classical Nambu dynamics \cite{nambu1973generalized}.) This explains the persistence of
positive energy states even for strong dephasing. We believe that the latter observation should also hold
in condensed matter physics. 

\section{Conclusions} 
\label{Sec:conclusions}

We introduced the density matrix formalism for relativistic quantum
mechanics as a generalization of the spinorial description of the Dirac
equation. This formalism is employed to describe interactions with an environment. Moreover,
we presented concise and effective numerical algorithms for the density matrix as well as the
relativistic Wigner function propagation.

As a particularly important case, a Lindbland model of quantum dephasing was studied.
While  decoherence eliminated interferences, the particular structure of a free
Majorana spinor remained robust. Partial robustness was also observed for a coordinate dependent mass term 
in the Dirac equation.
This robustness represents yet another remarkable attribute of Majorana
spinors \cite{RevModPhys.87.137} not presently acknowledged, which may be important 
experimentally. Moreover, the dynamics of the Klein paradox as well as Klein tunneling 
turned out to be weakly affected by quantum dephasing.

The presented numerical approach opens new horizons in a number of fields such as relativistic 
quantum chaos \cite{tomaschitz1991relativistic}, the quantum-to-classical transition, and experimentally 
inspired relativistic atomic and molecular physics \cite{RevModPhys.78.309, RevModPhys.81.163, PhysRevLett.110.255002}. 
Additionally, our method can be used to simulate effective systems modeled by relativistic 
mechanics, e.g., graphene \cite{PhysRevLett.103.025301,morandi2011wigner}, trapped ions \cite{gerritsma2010quantum}, 
optical lattices \cite{PhysRevLett.105.143902}, 
and semiconductors \cite{PhysRevLett.94.206801,zawadzki2011zitterbewegung}. 
Finally, the developed techniques can be generalized to treat 
Abelian   \cite{vasak1987quantum,mendoncca2011wave,PhysRevE.85.056411} as well as 
non-Abelian \cite{elze1986transport,elze1989quark} (e.g., quark gluon) plasmas.

\emph{Acknowledgments.} 
The authors thank Wojciech Zurek for insightful comments. R.C. is supported by DOE DE-FG02-02ER15344, D.I.B., and H.A.R. are partially supported by ARO-MURI W911NF-11-1-0268. A.C. acknowledges the support of the Fulbright Program. D.I.B. was also supported by 2016 AFOSR
Young Investigator Research Program.

%\input{RelativisticWigner.bbl}

%\end{thebibliography}

\newpage

\appendix

\section{Lorentz covariance of the Dirac equation}\label{Sec:LorentzCovariance}

A vector in Feynman's slash notation reads 
\begin{align}
  u\!\!\!/ = u^\mu \gamma_\mu,
\end{align}
where the gamma matrices obey the following Clifford algebra
\begin{align}
 \gamma_\mu \gamma_\nu + \gamma_\nu \gamma_\mu = 2  g_{\mu \nu} \mathbf{1},
\label{Cl4}
\end{align}
with $g_{\mu \nu} = diag(1,-1,-1,-1)$. The restricted Lorentz transform
does not carry out reflections and  preserves the direction of time and belongs to the group referred as $SO_{+}(1,3)$. 
In the present case the transformation for the vector $ u\!\!\!/$  is carried out in terms of 
Lorentz spinors $L$  belonging to the double cover group of  $SO_{+}(1,3)$, according to 
\begin{align}
   u\!\!\!/ \rightarrow u^\prime\!\!\!\!\!/ = L u\!\!\!/ L^{-1}.
\label{Lorentz-u}
\end{align}
The concept of a spinor as an operator can be found for example in chapter 10 of Ref. \cite{lounesto2001clifford}.
The double cover of  $SO_{+}(1,3)$ is known as the  $\bold{Spin}_{+}(1,3)$ group and is precisely  defined as
\begin{align}
 \bold{Spin}_{+}(1,3) = \{ L \in Matrices(4,\mathbb{C}) |\,  L \gamma^0 L^{\dagger} \gamma^0 = \boldsymbol{1}  \}
\end{align} 
For this type of Lorentz transform the inverse can be obtained as \cite{lounesto2001clifford}
\begin{align}
L^{-1} = \gamma^{0}L^{\dagger}\gamma^0.
\end{align}
The restricted Lorentz transform can also be carried out by the action of
the complex special linear group $SL(2,\mathbb{C}) \simeq
\bold{Spin}_{+}(1,3) $
\cite{PhysRevA.45.4293,Baylis1996book,lounesto2001clifford}, which is
made of $2\times 2$ complex matrices with determinant one. The proper
orthochronous Lorentz transformations can be parametrized by 6
variables denoting rotations and boots
\begin{align}
 L = \exp{\left(\frac{1}{2} \eta_{k} \gamma^{0}\gamma^{k}  \right)}  
 \exp{\left( \frac{1}{4} \epsilon_{jkl} \theta^{j} \gamma^{k}\gamma^{l}  \right)}, 
\label{classical-Lorentz-rotor} 
\end{align}
where $\theta^j$ represent three rotation angles, $\eta_k$ three boosts (rapidity variables)
and $\gamma^\mu = \gamma_{\mu}^{-1}$. The proper velocity can be obtained as the active boost of the proper velocity of a 
particle initially at rest with proper velocity  $ u\!\!\!/_{rest} =  \gamma^0$. This means that in general 
it is possible to find a Lorentz spinor $L$ such that
\begin{align}
  u\!\!\!/ =  L  u\!\!\!/_{rest}  L^{-1} = L L^\dagger \gamma^0. 
     \label{general-L}
\end{align}
This expression indicates that the information stored in the 4-vector $u\!\!\!/$ 
can be carried out by the associated Lorentz rotor $L$ and the fixed reference 4-vector $u\!\!\!/_{rest}$.

The Lorentz transformation in Eq. (\ref{Lorentz-u}) implies that
\begin{align}
 \bar{u}_{\mu} \gamma^{\mu} = L u_{\mu} \gamma^{\mu} L^{-1}. 
\end{align}
Considering that $u_{\mu}$ transforms as the components of a covariant tensor, we obtain
 \begin{align}
 u_{\nu} \frac{\partial x^\nu}{\partial x^{\prime \mu} }  \gamma^{\mu} = u_{\nu} L  \gamma^{\nu} L^{-1}, 
\end{align}
which implies that
\begin{align}
 L \gamma^{\nu} L^{-1} =   \frac{\partial x^\nu}{\partial x^{\prime \mu} } \gamma^{\mu}.
\label{Lorentz-gamma}
\end{align} 

The Lorentz transformation of a vector field that depends on the spacetime position $x$ 
is carried out in a similar manner as (\ref{Lorentz-u})
\begin{align}
 A( x ) \rightarrow \bar{A}( \bar{x}) =
 L A( x )  L^{-1}. 
\end{align}
Moreover, assuming that the origins of the reference frames coincide,
\begin{align}
 \bar{A}(\bar{x}) =
 L A( L^{-1}  \bar{x} L  )  L^{-1}. 
\end{align}

The Lorentz transformation of a spinorial field is consistent accordingly 
\begin{align}
  \psi(x) \rightarrow 
   \bar{\psi}( \bar{x}) = L  \psi(x)
\end{align}

The manifestly covariant Dirac equation is
\begin{align}
  i c \hbar \gamma^\mu \frac{\partial}{\partial x^\mu} \psi(x)
 - \gamma^{\mu} e A_{\mu}(x)\psi(x) - mc^2 \psi(x)=0,
\end{align}
such that applying the Lorentz rotor $L$ on the left we obtain
\begin{align}
  i c \hbar L \gamma^\mu \frac{\partial}{\partial x^\mu} L^{-1} L \psi(x) 
- L \gamma^{\mu} e A_{\mu}(x) L^{-1} L \psi(x) -  mc^2 L \psi(x) = 0,
\end{align}
Employing Eq. (\ref{Lorentz-gamma}), the first term of this equation 
can be written as 
\begin{align}
    i \hbar L \gamma^\mu \frac{\partial}{\partial x^\mu} L^{-1} L \psi(x) =&
   i \hbar  \frac{\partial x^\mu}{\partial \bar{x}^{ \nu} } \gamma^{\nu}   \frac{\partial}{\partial x^\mu} \bar{\psi}( \bar{x}) \\
   =& i \hbar \gamma^{\nu}  \frac{\partial}{\partial \bar{x}^{\nu}} \bar{ \psi}( \bar{x}).      
\end{align}
Therefore, maintaining the form for the Dirac equation and demonstrating its relativistic covariance
\begin{align}
  i \hbar \gamma^\mu \frac{\partial}{\partial \bar{x}^{ \mu}} \bar{\psi}(\bar{x})
 - \gamma^{\mu} e \bar{A}_{\mu}(\bar{x})\bar{\psi}(\bar{x}) =  mc \bar{\psi}(\bar{x}).
\end{align}
Furthermore, it follows that the relativistic density matrix $P(x,x^\prime) = \psi(x)\psi^\dagger(x^\prime)\gamma^0$  transforms as
\begin{align}
 P(x,x^\prime) \rightarrow \bar{P}( \bar{x}, \bar{x}^\prime )
&= \bar{\psi}(\bar{x}) \bar{\psi}^\dagger( \bar{x}^\prime)\gamma^0 \\ 
&= L  \psi(x)\psi^\dagger(x^\prime) L^\dagger \gamma^0 \\
&=  L  \psi(x)\psi^\dagger(x^\prime)\gamma^0 \gamma^0 L^\dagger \gamma^0\\
&= L P(x,x^\prime) L^{-1}.
\end{align}

\section{The classical limit of the Dirac equation}\label{Sec:ClassLimitDirac}
 
 The Dirac equation reads
\begin{align}
	D\psi=\left[\gamma^0\gamma^\mu(c\hat{\boldsymbol{p}}_\mu-e A_\mu(\hat{\boldsymbol{x}}))-\gamma^0m c^2\right]\psi=0. \label{dirac1}
\end{align}
In the classical limit, we understand the situation when the operators of the momenta  $\hat{p}_\mu$ and coordinates $\hat{x}^{\mu}$ commute \cite{Dirac1958, Shirokov1979d, PhysRevLett.109.190403}. Following the Hilbert phase space formalism \cite{PhysRevLett.109.190403, bondar2012wigner}, we separate the commutative and non-commutative parts of the Dirac generator $D$ by introducing the algebra of classical observables 
\begin{align}
[\hat{x}^\mu,\hat{p}_\nu]=0,\qquad
[\hat{p}_\mu,\hat{\theta}^\nu]=-i\delta^\nu_\mu,\\
[\hat{x}^\mu,\hat{\lambda}_\nu]=-i\delta^\mu_\nu, \qquad
[\hat{\lambda}_\mu,\hat{\theta}^\nu]=0,
\end{align}
which is connected with the quantum observables as
\begin{align}
\hat{\boldsymbol{x}}^\mu=\hat{x}^\mu- \hbar \hat{\theta}^\mu / 2,\qquad
\hat{\boldsymbol{p}}_\mu=\hat{p}_\mu+ \hbar \hat{\lambda}_\mu /2 .
\label{subst}
\end{align}
Substituting Eq. (\ref{subst}) into Eq. (\ref{dirac1}) and keeping the
terms up to the zero-th order in $\hbar$, we get a function of
$\hat{x}^{\mu}$ and $\hat{p}_{\mu}$. Considering that $\hat{x}^{\mu}$
and $\hat{p}_{\mu}$ commute, we drop the hat hereafter such that
\begin{align}
D=\gamma^0\gamma^\mu( cp_\mu - e A_\mu)-\gamma^0mc^2 + O(\hbar).
\label{classical}
\end{align}
 Utilizing the following unitary operator $U$
\begin{align}
& U=  \sqrt{\frac{ E_p+mc^2}{2E_p}}  
 \left( \textbf{1} -  \frac{  \gamma^{k}  (c p_{k}-eA_{k})  }{E_p+mc^2}   \right),
\label{eq3} 
\end{align}
\begin{align}
E_p&=\sqrt{ (mc^2)^2  + ( cp-eA)^k \cdot (cp-eA )^k  },
\end{align}
we finally obtain 
\begin{align}
\label{diagonalDiracG}
\lim_{\hbar\to0} UDU^\dagger  &= \left( 
	\begin{array}{cccc}
		H_{+} &	&	& 0\\
			& H_{+} &	& \\
			&	& H_{-} & \\
		0	&	& 	& H_{-} \\
	\end{array} \right),
\end{align}
with
\begin{align}
 H_{\pm} &=  cp_0 - e A_0 \pm E_p.
\label{classical_a}
\end{align}
According to Eq. (\ref{classical_a}), the Dirac generator $D$ in the classical limit corresponds to a decoupled pair of classical time-extended Hamiltonians. The Hamiltonian $H_{+}$ describes the dynamics of a classical relativistic particle; while, $H_{-}$ governs the dynamics of a particle traveling backwards in time, which resembles an antiparticle. These conclusions confirm the results of numerical simulations in the main text, where a Dirac particle was coupled to a bath causing decoherence that physically realizes the $\hbar\to 0$ limit.

%\bibliography{bib-relativity}

\end{document}